%% file: master.tex
\documentclass[11pt, a4paper]{article}

\usepackage{jheppub}
\usepackage{axodraw4j}
\usepackage{pstricks}
\usepackage{color}

\allowdisplaybreaks[4]

\input{macros}
\title{\boldmath The real radiation  antenna function for $S \to Q {\bar Q} q {\bar q}$ at
NNLO QCD}
\author{Werner Bernreuther,}
\author{Christian Bogner}
\author{and Oliver Dekkers}
\affiliation{Institut f\"ur Theoretische Teilchenphysik und Kosmologie,
RWTH Aachen University,\\
D-52056 Aachen, Germany}
\emailAdd{breuther@physik.rwth-aachen.de}
\emailAdd{bogner@physik.rwth-aachen.de}
\emailAdd{dekkers@physik.rwth-aachen.de}
\abstract{
As a first step towards the application of  the antenna subtraction formalism
 to NNLO QCD reactions with massive quarks, we 
 determine the real radiation  antenna function and its
 integrated counterpart for reactions of the type $S \to Q{\bar Q} q
 {\bar q}$, where $S$ denotes an uncolored initial state and
 $Q$, $q$ a massive and massless quark, respectively.
We compute the corresponding integrated antenna function 
   in terms of harmonic
  polylogarithms.  As an application and
   check of our results we calculate   the contribution
proportional to $\alpha_s^2 e^2_Q N_f$ to the inclusive heavy-quark
pair production cross section in $e^+e^-$ annihilation. 
}
\begin{document}
\maketitle 
%
%
\input{QQqq_Introduction}
\input{QQqq_Antenna_Subtraction}
\input{QQqq_Integrated_Antenna}

\input{QQqq_Check}
\input{QQqq_conclusion_acknowledgements}
\appendix
%
\input{QQqq_Appendix_Masters}

\input{QQqq_Bibliography}
\end{document}

%% file: macros.tex
\newcommand{\psmass}[1]
 { \!\frac{d^{d-1} {#1}}{ (2\pi)^{d -1} 2#1^0} \:}

\newcommand{\einhalb}{ \frac{1}{2} }

\newcommand{\cA}{\ensuremath{\mathcal{A}}}
\newcommand{\cB}{\ensuremath{\mathcal{B}}}

\newcommand{\cM}{\ensuremath{\mathcal{M}}}

\newcommand{\cO}{\ensuremath{\mathcal{O}}} 

\newcommand{\cT}{\ensuremath{\mathcal{T}}}

\newcommand{\mss}[1]{ {\mbox{\scriptsize #1}} }

\newcommand{\mycomment}[1]{ } 

%% file: QQqq_Introduction.tex
%
%
\section{Introduction}
\label{Intro}
The calculation of differential cross sections and distributions in
perturbative  QCD beyond the leading order requires, apart from the
renormalization of the QCD coupling and the quark masses, methods to
regularize and handle the infrared (IR) divergencies that appear in the 
intermediate steps of such computations. One general approach is to construct
subtraction terms such that, after adding/subtracting these terms, 
the IR singularities are regulated and cancelled
in tree-amplitudes involving the radiation of real massless partons and  in
associated loop amplitudes in the computation of IR safe observables (up to
factorization of collinear initial-state singularities).

For calculations at NLO QCD a widely used version of this approach is the
dipole subtraction method for massless QCD  \cite{Catani:1996vz} and for QCD
with massive quarks and other colored massive particles
\cite{Catani:2000ef,Phaf:2001gc,Catani:2002hc}, which was slightly
modified in \cite{Nagy:1998bb,Campbell:2004ch,Czakon:2009ss,Bevilacqua:2009zn,Frederix:2010cj}
and has found a number of computer implementations 
\cite{Gleisberg:2007md,Seymour:2008mu,Hasegawa:2009tx,Czakon:2009ss,
Frederix:2010cj}.
Other NLO    subtraction methods were constructed and applied, too,
including those of 
\cite{Frixione:1995ms,Nagy:1996bz,Frixione:1997np,Chung:2010fx}, and
  the antenna method (see below).

Computations of differential cross sections at NNLO QCD involve three
types of contributions: squares of tree-level double real emission amplitudes
(with $n+2$ final-state partons), 
interferences of  one-loop and
  tree-level amplitudes ($n+1$ final-state
partons) and of Born, one-loop, and two-loop
amplitudes ($n$ final-state
partons). For general discussions of the IR structure, see
\cite{Campbell:1997hg,Catani:1999ss,Kosower:2002su}.
  Various techniques have been devised  to handle
 the IR divergences of these individual contributions. These
 include the sector decomposition algorithm
\cite{Binoth:2003ak,Binoth:2004jv,Anastasiou:2003gr}, the antenna formalism 
 \cite{Kosower:1997zr,Kosower:2003bh,GRGG1}, and the subtraction methods
\cite{Weinzierl:2003fx,Weinzierl:2003ra,Frixione:2004is,Somogyi:2006da,Somogyi:2008fc,Bolzoni:2009ye,Catani:2007vq,Czakon:2010td,Czakon:2011ve}.
Application to  reactions at NNLO QCD include  $pp \to H + X$
\cite{Anastasiou:2004xq,Catani:2007vq},  $pp \to W + X$ \cite{Melnikov:2006di,Catani:2009sm},
$e^+ e^- \to 2 \ {\rm
      jets}$   \cite{Anastasiou:2004qd,GehrmannDeRidder:2004tv}, 
and $e^+ e^- \to 3 \ {\rm jets}$
 \cite{GehrmannDeRidder:2007jk,GehrmannDeRidder:2008ug,Weinzierl:2008iv,
Weinzierl:2009nz}, where the jet calculations
   just mentioned were made for massless partons. 

The antenna method  \cite{Kosower:1997zr,Kosower:2003bh,GRGG1}, which
we employ in this paper,  was
first worked out fully to NNLO QCD for $e^+e^-$ annihilation into  massless 
final-state
partons. The general set-up of this approach applies also to colored
initial
states and/or massive colored particles in the final state. The
 extension to 
processes with  initial-state and massless final-state partons  at NLO QCD
 was made in \cite{Kosower:1997zr,Kosower:2003bh,Daleo:2006xa}. For
  reactions with initial-state and massless final-state partons  at
  NNLO QCD, results were presented in
  \cite{Daleo:2009yj,Glover:2010im,Boughezal:2010mc}. For processes with massive
  quarks  $Q$ in the final state, the antenna
  subtraction terms at NLO QCD were explicitly worked out for colorless initial
states
  and for the hadronic reactions $h_1 h_2 \to Q {\bar Q}, \  Q {\bar Q}
    + \ {\rm jet}$ in \cite{GehrmannDeRidder:2009fz} and in
    \cite{Abelof:2011jv}, respectively.
 As is well-known,  the IR singularity structure of the matrix elements for
 reactions with massive colored particles is less entangled
 than that of their massless counterparts but, on the other hand, the
  (analytical) computation of the integrated subtraction terms is more
  difficult. 

In this paper we are concerned with  the production of a heavy-quark pair $Q
{\bar Q}$ by an uncolored initial state $S$ at NNLO QCD, i.e., we consider 
reactions of the type
\begin{equation} \label{intro:eq1}
 S \to Q \ {\bar Q} \ + X \, ,
\end{equation}
at order $\alpha_s^2$. This includes the production of a pair of heavy quarks
by electron-positron annihilation,  $e^+e^- \to \gamma^*, Z^* \to  Q  {\bar
Q} X$,  and the decay of a color and electrically neutral massive boson of
any spin into $Q \bar{Q} X$. The following ingredients are neccessary for the
computation of
arbitrary differential distributions to order $\alpha_s^2$: i) The amplitudes
$S\to   Q {\bar Q}$ to order $\alpha_s^2$. They are known in analytical form
for $S =$ vector \cite{Bernreuther:2004ih,Gluza:2009yy}, axial vector \cite{Bernreuther:2004th,Bernreuther:2005rw},
 scalar and pseudoscalar  \cite{Bernreuther:2005gw}. ii)
    The tree- and one-loop amplitudes for $S \to   Q {\bar   Q} g$. The
    one-loop amplitudes  can be
    computed with standard methods and are known for $e^+e^-$
    annihilation, 
  i.e. $S=\gamma^*, Z^*$ \cite{Brandenburg:1997pu,Nason:1997nw,Rodrigo:1999qg}.
iii) The tree-level
    amplitudes $S\to  Q {\bar Q} Q {\bar Q},$  $Q {\bar Q} gg,$ and $Q {\bar
      Q} q {\bar q}$,  where $q$ denotes a massless quark.   
     Apart from the tree-level amplitudes $S\to Q  {\bar
   Q}$ and $ S\to  Q {\bar Q} Q {\bar Q}$,
   the matrix elements   give rise to IR singularities, which are
  regulated within the above-mentioned subtraction methods by
  appropriate counterterms.
   The calculation of the differential cross
  section $d\sigma_{\rm NLO}$ for
 $S$ decaying into two massive quark jets is standard.
  The contribution of order $\alpha_s^2$ to the two-jet cross section
  is given schematically, using the
  notation of \cite{GehrmannDeRidder:2004tv},   by 
\begin{align}
d\sigma_{\rm NNLO}  &  = \int_{d\Phi_4} \left(d\sigma_{\rm NNLO}^R -
    d\sigma_{\rm NNLO}^S\right)
+ \int_{d\Phi_4} d\sigma_{\rm NNLO}^S  \nonumber \\
& + \int_{d\Phi_3} \left(d\sigma_{\rm NNLO}^{V,1} -
    d\sigma_{\rm NNLO}^{VS,1} \right)
  +   \int_{d\Phi_3} d\sigma_{\rm NNLO}^{VS,1}  +   \int_{d\Phi_2}
  d\sigma_{\rm NNLO}^{V,2} \, . \label{intro:eq2}
\end{align}
Here $d\sigma_{\rm NNLO}^R$, $d\sigma_{\rm NNLO}^{V,1}$, and
$d\sigma_{\rm NNLO}^{V,2}$ denote the contributions from the 
 tree-level amplitudes  $Q {\bar Q} gg$ and $Q {\bar Q} q {\bar q}$
(and $Q {\bar Q} Q {\bar Q}$, which does not require a subtraction
term), the one-loop  $Q {\bar Q} g$ amplitude, and the two-loop   $Q
 {\bar Q}$ amplitude, respectively. The term  $d\sigma_{\rm NNLO}^S$
  $(d\sigma_{\rm NNLO}^{VS,1})$ is a subtraction term 
 that coincides with $d\sigma_{\rm NNLO}^R$ $(d\sigma_{\rm
   NNLO}^{V,1})$ in all singular limits. 

The IR singularities of the two-loop term  $d\sigma_{\rm NNLO}^{V,2}$
  are explicitly known within dimensional 
regularization
\cite{Bernreuther:2004ih,Gluza:2009yy,Bernreuther:2004th,Bernreuther:2005rw,
Bernreuther:2005gw}.
The construction of  $d\sigma_{\rm NNLO}^S$
and $d\sigma_{\rm NNLO}^{VS,1}$ depends on the subtraction method used
 -- the integration of these terms over the four- and three-parton
 phase spaces $d\Phi_4$ and $d\Phi_3$, respectively, is in any case a
 difficult task. To our knowledge this has not yet been done for
 massive quarks in analytical form. As mentioned above, we shall use the antenna
framework. As a first step in the computation of $d\sigma(Q{\bar Q})$ to NNLO
QCD within this approach, we determine in this paper the subtraction term for
the $Q {\bar Q} q {\bar q}$ final state and its integral over the four-parton
phase space. The new aspect of this computation is the analytic integration
of the massive tree-level antenna function associated with the process
\begin{equation} \label{intro:eq3}
 \gamma^*(q) \to Q(p_1) \ {\bar Q}(p_2) \ + q(p_3)\ {\bar q}(p_4) \, .
\end{equation}
over the full four-particle phase space. This (integrated) antenna
function is not only of relevance for the specific process at hand, but serves
also as a building block for constructing subtraction terms for other processes
\eqref{intro:eq1} within the antenna formalism. 

The paper is organized as follows. In Section 2 we determine the
antenna function for  (\ref{intro:eq3}) and in Section 3 we integrate
this function analytically over the four-parton phase space. As an
application and check of our results, we compute in Section 4 the cross
section for the inclusive production of a massive $Q{\bar Q}$ pair
plus $N_f$ massless quarks by $e^+e^-$ annihilation via a virtual
photon -- more precisely, the contribution of order
$e_Q^2\alpha_s^2N_f$ to this cross section. Section 5 contains a
summary and outlook.

%

%% file: QQqq_Antenna_Subtraction.tex
%
%
\section{Antenna subtraction at NNLO QCD}
\begin{figure}
\centerline{\resizebox{0.6\linewidth}{!}{ \fcolorbox{white}{white}{
  \begin{picture}(319,329) (10,-12)
    \SetWidth{3.0}
    \SetColor{Black}
    \Line[arrow,arrowpos=0.5,arrowlength=8.333,arrowwidth=3.333,arrowinset=0.2,flip](61.11,256.193)(108.118,206.835)
    \SetWidth{1.0}
    \Gluon(84.614,280.872)(112.819,280.872){-2.938}{3}
    \Line[arrow,arrowpos=0.5,arrowlength=5,arrowwidth=2,arrowinset=0.2](112.819,280.872)(131.622,297.325)
    \Line[arrow,arrowpos=0.5,arrowlength=5,arrowwidth=2,arrowinset=0.2,flip](112.819,280.872)(131.622,264.419)
    \SetWidth{3.0}
    \Line[arrow,arrowpos=0.5,arrowlength=8.333,arrowwidth=3.333,arrowinset=0.2](61.11,256.193)(84.614,280.872)
    \Text(9.402,264.419)[lb]{\large{\Black{$\gamma^\ast$}}}
    \Text(150.425,297.325)[r]{\Large{\Black{$p_3$}}}
    \Text(150.425,264.419)[r]{\Large{\Black{$p_4$}}}
    \Text(126.921,206.835)[rt]{\Large{\Black{$p_2$}}}
    \Text(126.921,305.551)[rb]{\Large{\Black{$p_1$}}}
    \SetWidth{1.0}
    \Line[dash,dashsize=5.876](192.732,256.193)(239.74,256.193)
    \SetWidth{3.0}
    \Line[arrow,arrowpos=0.5,arrowlength=8.333,arrowwidth=3.333,arrowinset=0.2](239.74,256.193)(286.748,305.551)
    \Line[arrow,arrowpos=0.5,arrowlength=8.333,arrowwidth=3.333,arrowinset=0.2,flip](239.74,256.193)(263.244,231.514)
    \Line[arrow,arrowpos=0.5,arrowlength=8.333,arrowwidth=3.333,arrowinset=0.2,flip](263.244,231.514)(286.748,206.835)
    \SetWidth{1.0}
    \Gluon(263.244,231.514)(291.449,231.514){2.938}{3}
    \Line[arrow,arrowpos=0.5,arrowlength=5,arrowwidth=2,arrowinset=0.2](291.449,231.514)(310.252,247.967)
    \Line[arrow,arrowpos=0.5,arrowlength=5,arrowwidth=2,arrowinset=0.2,flip](291.449,231.514)(310.252,215.061)
    \SetWidth{3.0}
    \Line[arrow,arrowpos=0.5,arrowlength=8.333,arrowwidth=3.333,arrowinset=0.2](84.614,280.872)(108.118,305.551)
    \SetWidth{1.0}
    \Line[dash,dashsize=5.876](14.102,256.193)(61.11,256.193)
    \Line[dash,dashsize=5.876](14.102,75.213)(61.11,75.213)
    \Line[arrow,arrowpos=0.5,arrowlength=5,arrowwidth=2,arrowinset=0.2](61.11,75.213)(84.614,99.892)
    \Line[arrow,arrowpos=0.5,arrowlength=5,arrowwidth=2,arrowinset=0.2,flip](61.11,75.213)(108.118,25.854)
    \Line[arrow,arrowpos=0.5,arrowlength=5,arrowwidth=2,arrowinset=0.2](84.614,99.892)(108.118,124.571)
    \Gluon(84.614,99.892)(112.819,99.892){-2.938}{3}
    \SetWidth{3.0}
    \Line[arrow,arrowpos=0.5,arrowlength=8.333,arrowwidth=3.333,arrowinset=0.2](112.819,99.892)(131.622,116.344)
    \Line[arrow,arrowpos=0.5,arrowlength=8.333,arrowwidth=3.333,arrowinset=0.2,flip](112.819,99.892)(131.622,83.439)
    \SetWidth{1.0}
    \Line[dash,dashsize=5.876](192.732,75.213)(239.74,75.213)
    \Line[arrow,arrowpos=0.5,arrowlength=5,arrowwidth=2,arrowinset=0.2](239.74,75.213)(286.748,124.571)
    \Line[arrow,arrowpos=0.5,arrowlength=5,arrowwidth=2,arrowinset=0.2,flip](239.74,75.213)(263.244,50.533)
    \Gluon(263.244,50.533)(291.449,50.533){2.938}{3}
    \Line[arrow,arrowpos=0.5,arrowlength=5,arrowwidth=2,arrowinset=0.2,flip](263.244,50.533)(286.748,25.854)
    \SetWidth{3.0}
    \Line[arrow,arrowpos=0.5,arrowlength=8.333,arrowwidth=3.333,arrowinset=0.2](291.449,50.533)(310.252,66.986)
    \Line[arrow,arrowpos=0.5,arrowlength=8.333,arrowwidth=3.333,arrowinset=0.2,flip](291.449,50.533)(310.252,34.081)
    \Text(75.213,165.703)[lb]{\Large{\Black{$(C_1)$}}}
    \Text(253.842,165.703)[lb]{\Large{\Black{$(C_2)$}}}
    \Text(75.213,-15.278)[lb]{\Large{\Black{$(C_3)$}}}
    \Text(253.842,-15.278)[lb]{\Large{\Black{$(C_4)$}}}
    \Text(188.031,264.419)[lb]{\large{\Black{$\gamma^\ast$}}}
    \Text(9.402,83.439)[lb]{\large{\Black{$\gamma^\ast$}}}
    \Text(188.031,83.439)[lb]{\large{\Black{$\gamma^\ast$}}}
    \Text(150.425,83.439)[r]{\Large{\Black{$p_2$}}}
    \Text(329.055,34.081)[r]{\Large{\Black{$p_2$}}}
    \Text(126.921,25.854)[rt]{\Large{\Black{$p_4$}}}
    \Text(126.921,124.571)[rb]{\Large{\Black{$p_3$}}}
    \Text(150.425,116.344)[r]{\Large{\Black{$p_1$}}}
    \Text(329.055,66.986)[r]{\Large{\Black{$p_1$}}}
    \Text(305.551,124.571)[rb]{\Large{\Black{$p_3$}}}
    \Text(305.551,25.854)[rt]{\Large{\Black{$p_4$}}}
    \Text(305.551,206.835)[rt]{\Large{\Black{$p_2$}}}
    \Text(305.551,305.551)[rb]{\Large{\Black{$p_1$}}}
    \Text(329.055,247.967)[r]{\Large{\Black{$p_3$}}}
    \Text(329.055,215.061)[r]{\Large{\Black{$p_4$}}}
  \end{picture}
} } }
\caption{Feynman diagrams contributing to $\gamma^\ast \to Q \bar{Q} q
\bar{q}$ at tree-level. Bold (thin) lines refer to massive (massless) quarks.}
\label{QQA_ABB_1}
\end{figure}
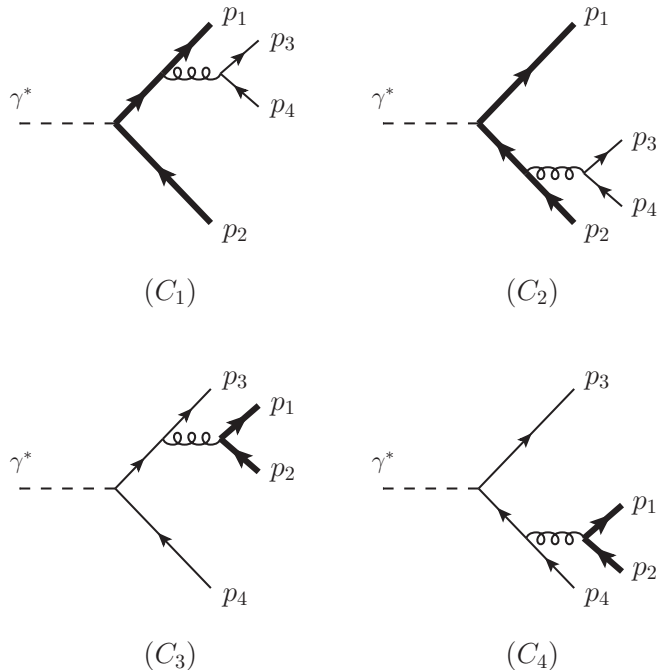

In the following, we restrict our attention to the case of reaction
\eqref{intro:eq3} where, for definiteness, we
 consider one massless quark flavor $q$ in the final state.
   As mentioned above, we focus on constructing a subtraction
term which coincides with the squared matrix element of $\gamma^\ast \to Q
\bar{Q} q \bar{q}$ in all single and double unresolved limits. In
fig.\,\ref{QQA_ABB_1}
   the Feynman diagrams are shown that are associated with this process at order $\alpha_s^2$.
   The corresponding
contribution to the cross section for 2-jet production may be written as follows:
\begin{eqnarray}
d \sigma^{R, Q \bar{Q} q \bar{q}}_\mss{NNLO} & = & 4 \pi
\alpha \left( 4 \pi \alpha_s \right)^2 \left( N_c^2 - 1 \right) d \Phi_4 ( p_1,
p_2, p_3, p_4; q )\,  J^{(4)}_2 ( p_1, p_2, p_3, p_4 )
\nonumber \\[0.2cm]
& & {} \times \left\{ e^2_Q \left| \cM^{0}_{Qq\bar{q}\bar{Q}} \right|^2 +
e^2_q \left| \cM^{0}_{qQ\bar{Q}\bar{q}} \right|^2 + 2 e_Q e_q\,
\mbox{Re}\left( \cM^{0}_{Qq\bar{q}\bar{Q}}\,
\cM^{0,\dagger}_{qQ\bar{Q}\bar{q}} \right) \right\} ,
\label{QQA_3}
\end{eqnarray}
where the matrix elements $\cM^{0}_{Q q \bar{q} \bar{Q}}$ 
 and $\cM^{0}_{q Q \bar{Q} \bar{q}}$
correspond to the
diagrams $C_1$, $C_2$ and $C_3$, $C_4$, respectively. 
 The
dependence on the electromagnetic and strong coupling and the
dependence on the number of colors $N_c$ are extracted
from the matrix elements; $e_Q$ ($e_q$) is the electric charge of the
massive
 (massless) quark in units of the positron charge.
 Summation over all spins is understood. In the formulae below, the
 polarizations of $\gamma^*$ are summed, but not averaged. 

   The  phase space measure $d \Phi_4$  in $d = 4 - 2 \epsilon$
   dimensions is
\begin{equation} \label{gm_3}
d \Phi_4 ( p_1, p_2, p_3, p_4 ; q ) = \mu^{ 12 - 3d } \prod^{4}_{i = 1}
\: \psmass{p_i}\! \left( 2 \pi \right)^{d}
\delta^{(d)}\! \left( q - \sum^4_{i = 1} p_i  \right), 
\end{equation}
where $\mu$ is a mass scale.
 The jet function
$J^{(n)}_m$  in (\ref{QQA_3})  ensures that only configurations are taken into
account where $n$ outgoing partons
form $m$ jets.  

Because the squared matrix element $\left| \cM^0_{q Q \bar{Q}
\bar{q}} \right|^2$  does not involve  infrared singular configurations,
no subtraction is required. Its contribution to the $Q{\bar Q}$
production cross section is given in \cite{Wetzel:1981uc}.
  The same holds for the
interference terms between $\cM^0_{Q q \bar{q} \bar{Q}}$ and $\cM^0_{q Q \bar{Q}
\bar{q}}$. Moreover, due to Furry's theorem these terms yield a vanishing
contribution to the cross section if the observable under
consideration does not distinguish between quarks and antiquarks.

In the framework of antenna subtraction the main building blocks for
constructing NNLO subtraction terms are the antenna functions, which
can be derived from physical color-ordered squared matrix elements for
tree-level $1 \to 3$ and $1 \to 4$ processes and one-loop $1 \to 3$ processes.
A detailed and completely general
analysis of how the subtraction terms are constructed from the various antenna
functions is  given in \cite{GRGG1} for  massless final state partons.
  This procedure applies also to the case of massive quarks. Its application to the
specific process $\gamma^\ast \to Q \bar{Q} q \bar{q}$ is outlined
below.  The  three-parton tree-level antenna functions with
massive quarks have been
calculated in \cite{GehrmannDeRidder:2009fz,Abelof:2011jv}, whereas the 
four-parton tree-level and three-parton one-loop antenna functions 
   that involve massive quarks are still
missing. 

In the case of $\gamma^\ast \to Q \bar{Q} q \bar{q}$ the construction of the
subtraction terms may be divided into two parts. In a first step, the single
unresolved configurations are subtracted. Within the antenna method, the
corresponding subtraction term reads
\begin{eqnarray}
d \sigma^{S,a}_\mss{NNLO} & = & 4 \pi \alpha \left( 4 \pi \alpha_s \right)^2
e_Q^2 \left( N_c^2 - 1 \right) d \Phi_4 ( p_1, p_2, p_3, p_4; q )
\left| \cM^0_{Q \bar{Q}}  \right|^2
\nonumber \\
& & {} \times \bigg[ \, E^0_3\! \left( 1_Q, 3_q, 4_{\bar{q}} \right) 
A^{0}_3  \! \left( \widetilde{ (13) }_Q, \widetilde{ (43) }_g,
2_{\bar{Q}} \right) J^{(3)}_2 \!  \left( \widetilde{ p_{13} },
\widetilde{ p_{43} }, p_2 \right) 
\nonumber \\
& & \qquad {} + E^0_3\! \left(2_{\bar{Q}} 3_q, 4_{\bar{q}} \right)
A^{0}_3 \! \left( 1_Q , \widetilde{ (34) }_g, \widetilde{
(24)}_{\bar{Q}} \right) J^{(3)}_2 \! \left(  p_1, \widetilde{ p_{34}
}, \widetilde{ p_{24} } \right) \bigg].
\end{eqnarray}
The massive quark-antiquark antenna function $A^0_3 \! \left( i_Q, k_g,
j_{\bar{Q}} \right)$ and the quark-gluon antenna $E^0_3 \! \left( i_Q, j_q,
k_{\bar{q}} \right)$ with a massive radiator quark are given in 
\cite{GehrmannDeRidder:2009fz}. The momenta $\widetilde{ p_{ik} }$ and $\widetilde{ p_{jk}
}$ are redefined on-shell momenta, constructed from linear combinations of the
momenta $p_i$, $p_j$ and $p_k$.
 The tree-level
two-parton matrix element squared (summed over colors and
spins, with
the photon coupling and $N_c$ factored out)  is
\begin{equation} \label{MQQ}
\Big| \cM^0_{Q\bar{Q}} (\gamma^\ast \to Q
\bar{Q}) \Big|^2 = 4 \left[ \left( 1 - \epsilon
\right) q^2 + 2 m^2 \right] \, ,
\end{equation}
where $m$ denotes the mass of $Q$.
In a second step, the  double unresolved configuration, where both $q$ and
$\bar{q}$ become soft, has to be subtracted. In the case at hand,
  the appropriately normalized      squared matrix element $\left|
\cM^{0}_{Qq\bar{q}\bar{Q}} \right|^2$ can be used as subtraction term.
 In the terminology of  \cite{GRGG1}   this is
 the  antenna function $B^0_4 ( 1_Q, 3_q, 4_{\bar{q}}, 2_{\bar{Q}} )$ associated with
the color-ordering $Q q \bar{q} \bar{Q}$, where a color-connected massless
quark-antiquark pair is radiated between a pair of massive quarks. More
precisely, this antenna function is defined by
\begin{equation} \label{ma}
B^0_4 ( 1_Q, 3_q, 4_{\bar{q}}, 2_{\bar{Q}} ) = \frac{ \Big| \cM^{0}_{Q q \bar{q}
\bar{Q} } \Big|^2 }{ \Big| \cM^{0}_{Q\bar{Q}} \Big|^2} \,,
\end{equation}
where the normalization factor is given in \eqref{MQQ}. Obviously \eqref{ma} has
the appropriate behaviour in the singular double unresolved limit where
$q$ and
$\bar{q}$ become simultaneously soft. However \eqref{ma} contains
also singularities due to single unresolved limits, which have to be subtracted from
the antenna function. In the end, the corresponding subtraction term
for the double
unresolved configuration reads
\begin{eqnarray}
d \sigma^{S,b}_\mss{NNLO} & = & 4 \pi \alpha \left( 4 \pi \alpha_s \right)^2
e_Q^2
\left( N_c^2 - 1 \right) d \Phi_4 ( p_1, p_2, p_3, p_4; q )\, \left|
\cM^{0}_{Q\bar{Q}} \right|^2
\nonumber \\[0.1cm]
& & {} \times \bigg[\, B^0_4 ( 1_Q, 3_q, 4_{\bar{q}}, 4_Q ) - E^0_3\! \left(
1_Q, 3_q, 4_{\bar{q}} \right) A^0_3 \!
\left( \widetilde{(13)}_Q, \widetilde{(43)}_g, 2_{\bar{Q}} \right) 
\nonumber \\
& & \qquad {} - E^0_3\! \left(2_{\bar{Q}} 3_q, 4_{\bar{q}} \right) A^0_3 \!
\left( 1_Q, \widetilde{(34)}_g, \widetilde{(24)}_{\bar{Q}} \right) \bigg]
J^{(2)}_2 \! \left( \widetilde{ p_{134} }, \widetilde{ p_{243} }
\right) \, ,
\end{eqnarray}
where $\widetilde{ p_{ikl} }$ and $\widetilde{ p_{jkl} }$ are linear
combinations of the momenta $p_i$, $p_j$, $p_k$ and $p_l$. 

The subtracted differential cross section
\begin{equation}  \label{sbdifcrx}
 d \sigma^{R, Q \bar{Q} q \bar{q}}_\mss{NNLO} - d \sigma^{S,a}_\mss{NNLO} - d
\sigma^{S,b}_\mss{NNLO} 
\end{equation}
is free of IR divergences and can be integrated over the four-parton phase space
numerically in $d=4$ dimensions.

For the antenna function $B^0_4$ we find
\begin{eqnarray}
B^0_4 \! \left( 1_Q, 3_q, 4_{\bar{q}}, 2_{\bar{Q}} \right) & = & \frac{1}{
\left( q^2+ 2 m^2 \right) } \bigg\{ \; \frac{1}{ s_{34} s_{134}^2 } \left[
s_{12} s_{13} + s_{12} s_{14} + s_{13} s_{23} + s_{14} s_{24} \right] 
\nonumber \\
& & {} + \frac{1}{ s_{34} s_{234}^2 } \left[ s_{12} s_{23} + s_{12} s_{24} +
s_{13} s_{23} + s_{14} s_{24}  \right]
\nonumber \\
& & {} + \frac{1}{ s_{34}^2 s_{134}^2 } \big[ 2 s_{12} s_{13} s_{14} + s_{13}
s_{14} s_{24} + s_{13} s_{14} s_{23} \nonumber \\
& & {}  - s_{13}^2 s_{24} - s_{14}^2 s_{23} \big]
\nonumber \\
& & {} + \frac{1}{ s_{34}^2 s_{234}^2 } \big[ 2 s_{12} s_{23} s_{24} + s_{13}
s_{23} s_{24} + s_{14} s_{23} s_{24} 
\nonumber \\
& & {} - s_{13} s_{24}^2 - s_{14} s_{23}^2 \big]
\nonumber \\
& & {} + \frac{1}{s_{34} s_{134} s_{234} } \left[ 2 s_{12}^2 + s_{12} s_{23} +
s_{12} s_{24} + s_{12} s_{13} + s_{12} s_{14} \right]
\nonumber \\
& & {} + \frac{1}{ s_{34}^2 s_{134} s_{234} } \big[ - s_{13} s_{24}^2 - s_{14}
s_{23}^2 - s_{13}^2 s_{24} - s_{14}^2 s_{23} \nonumber \\[0.2cm]
& & {} + s_{13} s_{14} s_{23} + s_{13} s_{14} s_{24} + s_{13} s_{23} s_{24} +
s_{14} s_{23} s_{24}
\nonumber \\[0.2cm]
& & {} - 2 s_{12} s_{13} s_{24} - 2 s_{12} s_{14} s_{23} \big] +  \frac{ 2
s_{12} }{ s_{134} s_{234} } 
\nonumber \\
& & {} + m^2 \bigg( \frac{ 8 s_{13} s_{14} }{ s_{34}^2 s_{134}^2 } + \frac{ 8
s_{23} s_{24} }{ s_{34}^2 s_{234}^2 } - \frac{ 4 }{ s_{134}^2 }  - \frac{ 4 }{
s_{234}^2 } 
\nonumber \\
& & {} - \frac{ 2 }{ s_{34} s_{134}^2 } \left[ s_{12} + s_{23} + s_{24} \right]
- \frac{ 2 }{ s_{34} s_{234}^2 } \left[ s_{12} + s_{13} + s_{14} \right]
\nonumber \\
& & {} + \frac{ 2 }{ s_{34} s_{134} s_{234} } \left[ 4 s_{12} - s_{13} - s_{14}
- s_{24} - s_{23} \right] 
\nonumber \\
& & {} - \frac{ 8 }{ s_{34}^2 s_{134} s_{234} } \left[ s_{14} s_{23} +  s_{13}
s_{24} \right] \bigg) 
\nonumber \\
& & {} - m^4 \left( \frac{ 8 }{ s_{34}\, s_{134}^2 }  + \frac{ 8 }{ s_{34}\,
s_{234} ^2 } \right) \bigg\} + \cO(\varepsilon)\,, \label{QQA_5}
\end{eqnarray}
where  $s_{ij} = 2\, p_i\cdot p_j$ and $s_{ijk} = s_{ij} + s_{ik} +
s_{jk}$. For the sake of brevity we have not written down
 in (\ref{QQA_5}) the terms of order $\varepsilon$.
For the numerical computation of
    (\ref{sbdifcrx}) in $d=4$ dimensions, only the  four-dimensional
 antenna function $B^0_4$ is required. 
  However, the integrated antenna function, which we compute in the
  next section, must be determined with  $B^0_4$ in $d$ dimensions.

For completeness, we derive also the behaviour of the antenna function $B^0_4$ in the double soft limit
$p_3 \to \lambda p_3$, $p_4 \to \lambda p_4$ with $\lambda \to 0$.
We obtain $B^0_4 =  \lambda^{-4} S_{12} + {\cal O}(\lambda^{-3})$ with 
\begin{eqnarray}
 S_{12}( 3_q, 4_{\bar{q}} ) & = & \frac{2}{s^2_{34} \left( s_{13} + s_{14}
\right) \left( s_{23} + s_{24} \right) } \left( s_{12} s_{34} - s_{13} s_{24} -
s_{14} s_{23} \right) \nonumber \\
& & \qquad {} + \frac{2}{ s_{34}^2 } \left( \frac{ s_{13} s_{14} - s_{34} m^2
}{ \left( s_{13} + s_{14} \right)^2 } + \frac{ s_{23} s_{24} - s_{34} m^2 }{
\left( s_{23} + s_{24} \right)^2 } \right).
\label{eq:dousofac}
\end{eqnarray} 
The double-soft factor \eqref{eq:dousofac} is not identical to the 
 respective factor $S_{12}^{m=0}$ for a massless quark $Q$ given in
\cite{GRGG1}, but
 agreees with this factor in the limit $m\to 0$.
A respective difference  between the $m\neq 0$ versus $m=0$ case
 was shown in \cite{Czakon:2011ve}  for $Q{\bar Q} gg$
 final states
    in the double-soft gluon limit.

%% file: QQqq_Integrated_Antenna.tex
%
%
%
\section{The integrated \boldmath $Qq\bar{q}\bar{Q}$ antenna function}
The introduction of subtraction terms must be counterbalanced by 
 adding their integrated counterparts, cf. (\ref{intro:eq2}).
In the context of the antenna method this implies the analytic
integration of the antenna function $B^0_4
( 1_Q, 3_q, 4_{\bar{q}}, 2_{\bar{Q}})$ over the four-parton antenna
phase space $d \Phi_{X_{Q q \bar{q} \bar{Q}}}$ associated with a massive and a
massless quark-antiquark pair. The corresponding integrated antenna function
$\cB^0_{Q q \bar{q} \bar{Q}}$ is defined by 
\begin{equation} \label{gm_19}
\cB^0_{Q q \bar{q} \bar{Q}} ( q^2, m, \mu, \varepsilon) = \left( 8 \pi^2 \left(
4 \pi \right)^{-\varepsilon} e^{\varepsilon \gamma_E } \right)^2 \int \!
d\Phi_{X_{ Q q \bar{q} \bar{Q} }} \: B^0_4 ( 1_Q, 3_q, 4_{\bar{q}}, 2_{\bar{Q}}
)\,.
\end{equation}
The antenna phase space $d \Phi_{X_{Q q \bar{q} \bar{Q}}}$ is closely related to
the full four-particle phase space \eqref{gm_3}:
\begin{equation} \label{gm_18}
d\Phi_4 (p_1, p_2, p_3, p_4, q) = P_2(q^2, m )\, d\Phi_{X_{Q q \bar{q}
\bar{Q} }}, 
\end{equation}
where $P_2 ( q^2 , m )$ denotes the integrated phase space
for two particles of equal mass $m$ in $d = 4 - 2 \epsilon$
 space-time dimensions:
\begin{equation} \label{PS2}
P_2 ( q^2 , m ) = 2^{-3 + 2\epsilon }\, \pi^{ - 1 + \epsilon }
\, \frac{ \Gamma\! \left( 1 - \epsilon \right) }{ \Gamma\! \left( 2 - 2 \epsilon
\right) } \left( \frac{ \mu^2 }{ q^2 } \right)^{ \epsilon } \left( 1 -
\frac{4m^2}{q^2} \right)^{ \frac{1}{2} - \epsilon }.
\end{equation}
The integration of the antenna function over the antenna phase space has to be
performed analytically in $d = 4 - 2 \epsilon$ dimensions. 
\subsection{Reduction to master integrals}

In order to calculate the integrated antenna function \eqref{gm_19}, we first
reduce the number of integrals to be computed by exploiting linear dependences
between phase-space integrals with the help of integration-by-parts (IBP)
identities \cite{CheTka}. These  identities were
originally derived for loop integrals, but are also very useful in
 computing phase space integrals \cite{AnaMel,GehGehHei}. \\
Using the Cutkosky rules \cite{Cut} we introduce two massive and two
massless cut-propagators
\begin{eqnarray}
\frac{1}{D_i}& = & 2\pi i \delta ^+ \left( p^2_i-m^2 \right) =
\frac{1}{p^2_i-m^2+i0}-\frac{1}{p^2_i-m^2-i0}, \,\qquad i=1,\,2,\\
\frac{1}{D_i}& = & 2\pi i \delta ^+ \left( p^2_i \right) =
\frac{1}{p^2_i+i0}-\frac{1}{p^2_i-i0}, \,\qquad i=3,\,4.
\end{eqnarray}
Then we can write:
\begin{eqnarray}
d \phi _4 \left( p_1,\,p_2,\,p_3,\,p_4,\,q\right) & = & \frac{ \mu
^{12-3d} }{ i^4 (2\pi)^{3d} }\, \delta^{(d)}\! \left( q - \sum^4_{i = 1} p_i 
\right) \prod_{i=1}^4
\frac{d^d p_i}{D_i} .
\end{eqnarray}
Next  we express each of the invariants ${s_{ij},\,s_{klm}}$ in the
 $d$-dimensional version of \eqref{QQA_5}
in terms of the four cut-propagators and five further appropriately chosen
propagators and scalar products, such that each term of $\cB^0_{Q q \bar{q}
\bar{Q}} $ can be
viewed as a four-particle cut through a three-loop vacuum polarization diagram,
which may contain irreducible scalar products in the numerator. It was shown in
 \cite{AnaMel} that IBP reduction of such integrals can be carried out in the
same way as for loop integrals, using that integrals where at least
one of the four cut-propagators is missing in the integrand  do not contribute to the original
cut-integral. Using the implementation \texttt{AIR} \cite{AnaLaz}  of the Laporta
reduction algorithm \cite{Lap} we decompose the integral
\eqref{gm_19}
 along these lines. 
  As a result we can express  the integrated antenna function in terms
  of five independent integrals (master integrals):
\begin{eqnarray}
\cB^0_{Q q \bar{q} \bar{Q}} ( q^2, m, \mu, \epsilon) 
& = & \frac{ \left( 8 \pi^2 \left(
4 \pi \right)^{-\epsilon} e^{\epsilon \gamma_E } \right)^2 }{ P_2 ( q^2 ,
m ) \, | \cM_{Q \bar{Q} } |^2 } \, \Bigg\{ 
\Bigg[ \frac{64+208 z+48 z^2-10 z^3+5 z^4}{ z (1-z)^3 }\,
\frac{1}{\epsilon^2} 
\nonumber \\[0.2cm]
& & + \frac{ 2 \left( 32 - 448 z - 112 z^2 + 133 z^3 - 49 z^4 \right)}{ 3 z
( 1 - z )^3 } \, \frac{1}{\epsilon} 
\nonumber \\[0.2cm]
& & + \frac{2 \left(1720+5656 z-164 z^2-163 z^3+214 z^4\right)}{9 z (1-z)^3 }
+ \ensuremath{\mathcal{O}}( \epsilon ) \bigg]
\nonumber \\[0.2cm]
& & {} \times \left( q^2 \right)^{-1} T_1(q^2,m^2,\epsilon)
\nonumber \\[0.2cm]
& & + \Bigg[- \frac{4 \left(48+56 z+6 z^2-5 z^3\right)}{z (1 - z)^3 } \,
\frac{1}{\epsilon^2}
\nonumber \\[0.2cm]
& & + \frac{4 \left(72+56 z + 234 z^2  - 101 z^3\right)}{3 z (1 - z)^3 } \,
\frac{1}{\epsilon}
\nonumber \\[0.2cm]
& & - \frac{8 \left(1296+1222 z+153 z^2-241 z^3\right)}{9 z (1 - z)^3 }+
\ensuremath{\mathcal{O}} ( \epsilon ) \Bigg]
\nonumber \\[0.2cm]
& & {} \times \left( q^2 \right)^{-2} T_2 (q^2,m^2,\epsilon)
\nonumber \\[0.2cm]
 & & + \Bigg[ - \frac{4 \left(16 + 64 z + 26 z^2 - z^3\right)}{z (1 - z)^3 } \,
\frac{1}{\epsilon^2}
\nonumber \\[0.2cm]
& & - \frac{ 4 \left( 8 - 232 z - 50 z^2 + 13 z^3  \right) }{ 3 z \left(1 -
z\right)^3 }\, \frac{1}{\epsilon}
\nonumber \\[0.2cm]
& & - \frac{ 8 \left( 440 + 1730 z + 247 z^2 + 13 z^3 \right) }{ 9  z \left(
1  - z \right)^3 } + \ensuremath{\mathcal{O}}( \epsilon ) \Bigg]
\nonumber \\[0.2cm]
& & {} \times \left( q^2 \right)^{-2} T_3 (q^2,m^2,\epsilon)
\nonumber \\[0.2cm]
& & + \Bigg[ - \frac{8 \left(4 - z^2\right)}{3 (1-z) z } \frac{1}{ \epsilon } +
\frac{4 \left(20+12 z+49 z^2\right)}{9 (1-z) z} + \ensuremath{\mathcal{O}}(
\epsilon ) \Bigg]
\nonumber \\[0.2cm]
& & {} \times T_4( q^2, m^2 ,\epsilon)
\nonumber \\[0.2cm]
& & + \Bigg[ \frac{2 \left(4-z^2\right)}{3 (1-z) } \frac{1}{\epsilon } - \frac{4
\left(8+9 z+z^2\right)}{9 (1-z)}  + \ensuremath{\mathcal{O}}( \epsilon ) \Bigg]
\nonumber \\[0.2cm]
& & {} \times q^2 \, T_5(q^2, m^2, \epsilon) \Bigg\} \, , \label{Master}
\end{eqnarray}
where we introduced the dimensionless variable $z \equiv 4 m^2 / q^2$.
 The  five master
integrals are 		
\begin{eqnarray}
T_1 (q^2 , m^2, \epsilon ) & = & 
\parbox{0.15\linewidth}{
\resizebox{\linewidth}{!}{
\fcolorbox{white}{white}{
  \begin{picture}(194,162) (63,-31)
    \SetWidth{4.0}
    \SetColor{Black}
    \Arc(160,50)(64,180,540)
    \SetWidth{1.0}
    \Line(96,50)(64,50)
    \Line(224,50)(256,50)
    \Arc[clock](160,-23.333)(97.333,131.112,48.888)
    \Arc(160,123.333)(97.333,-131.112,-48.888)
    \Line[dash,dashsize=4.6](160,130)(160,-30)
  \end{picture}
}}}
= \int \! d\Phi_4(p_1, p_2, p_3, p_4; q)\,,
\label{Master1} \\[0.2cm]
T_2 (q^2 , m^2, \epsilon ) & = & \raisebox{2ex}{\makebox[5ex]{$s_{13}$}}
\hspace{ - 2ex }
\parbox{0.15\linewidth}{
\resizebox{\linewidth}{!}{
  \begin{picture}(194,162) (63,-31)
    \SetWidth{4.0}
    \SetColor{Black}
    \Arc(160,50)(64,180,540)
    \SetWidth{1.0}
    \Line(96,50)(64,50)
    \Line(224,50)(256,50)
    \Arc[clock](160,-23.333)(97.333,131.112,48.888)
    \Arc(160,123.333)(97.333,-131.112,-48.888)
    \Line[dash,dashsize=4.6](160,130)(160,-30)
  \end{picture}
}}
\mycomment{ \parbox{0.15\linewidth}{
\resizebox{\linewidth}{!}{
\fcolorbox{white}{white}{
  \begin{picture}(208,162) (49,-31)
    \SetWidth{4.0}
    \SetColor{Black}
    \Arc(160,50)(64,180,540)
    \SetWidth{1.0}
    \Line(96,50)(64,50)
    \Line(224,50)(256,50)
    \Arc[clock](160,-23.333)(97.333,131.112,48.888)
    \Arc(160,123.333)(97.333,-131.112,-48.888)
    \Line[dash,dashsize=4.6](160,130)(160,-30)
    \Text(64,66)[]{\Huge{\Black{$s_{13}$}}}
  \end{picture}
}}} }
= \int \! d\Phi_4(p_1, p_2, p_3, p_4; q) \: s_{13}\,, 
\label{Master2} \\[0.2cm]
T_3 (q^2 , m^2, \epsilon ) & = & \raisebox{2ex}{\makebox[5ex]{$s_{134}$}}
\hspace{ - 2ex }
\parbox{0.15\linewidth}{
\resizebox{\linewidth}{!}{
  \begin{picture}(194,162) (63,-31)
    \SetWidth{4.0}
    \SetColor{Black}
    \Arc(160,50)(64,180,540)
    \SetWidth{1.0}
    \Line(96,50)(64,50)
    \Line(224,50)(256,50)
    \Arc[clock](160,-23.333)(97.333,131.112,48.888)
    \Arc(160,123.333)(97.333,-131.112,-48.888)
    \Line[dash,dashsize=4.6](160,130)(160,-30)
  \end{picture}
}}
\mycomment{ \parbox{0.15\linewidth}{
\resizebox{\linewidth}{!}{
\fcolorbox{white}{white}{
  \begin{picture}(208,162) (49,-31)
    \SetWidth{4.0}
    \SetColor{Black}
    \Arc(160,50)(64,180,540)
    \SetWidth{1.0}
    \Line(96,50)(64,50)
    \Line(224,50)(256,50)
    \Arc[clock](160,-23.333)(97.333,131.112,48.888)
    \Arc(160,123.333)(97.333,-131.112,-48.888)
    \Line[dash,dashsize=4.6](160,130)(160,-30)
    \Text(64,66)[]{\Huge{\Black{$s_{134}$}}}
  \end{picture}
}}} }
= \int \! d\Phi_4(p_1, p_2, p_3, p_4; q) \:
s_{134}\,, 
\label{Master3} \\[0.2cm]
T_4 (q^2 , m^2, \epsilon ) & = & 
\parbox{0.15\linewidth}{
\resizebox{\linewidth}{!}{
\fcolorbox{white}{white}{
  \begin{picture}(194,162) (63,-31)
    \SetWidth{4.0}
    \SetColor{Black}
    \Arc(160,50)(64,180,540)
    \SetWidth{1.0}
    \Line(96,50)(64,50)
    \Line(224,50)(256,50)
    \Arc[clock](140.343,-23.835)(111.581,108.466,41.431)
    \Arc(185.901,145.585)(102.898,-141.834,-68.268)
    \Line[dash,dashsize=4.6](160,130)(160,-30)
  \end{picture}
}}}
= \int \! d\Phi_4(p_1, p_2, p_3, p_4; q) \:
\frac{1}{s_{134}}\,, 
\label{Master4} \\[0.2cm]
T_5 (q^2 , m^2, \epsilon ) & = & 
\parbox{0.15\linewidth}{
\resizebox{\linewidth}{!}{
\fcolorbox{white}{white}{
  \begin{picture}(194,162) (63,-31)
    \SetWidth{4.0}
    \SetColor{Black}
    \Arc(160,50)(64,180,540)
    \SetWidth{1.0}
    \Line(96,50)(64,50)
    \Line(224,50)(256,50)
    \Arc[clock](119.345,-19.876)(102.881,98.015,21.602)
    \Arc(200.655,119.876)(102.881,-158.398,-81.985)
    \Line[dash,dashsize=4.6](160,130)(160,-30)
  \end{picture}
}}}
 = \int \! d\Phi_4(p_1, p_2, p_3, p_4; q) \:
\frac{1}{s_{134} s_{234}} \,.
\label{Master5}
\end{eqnarray}
In these diagrammatic representations bold (thin) lines refer to massive
(massless) propagators.
In the case of $T_2$ and $T_3$, the invariants to the left of the cut-diagrams  denote
numerator factors. Note that $T_1$ is just the $d$-dimensional  phase space volume
associated with  two massless and two massive (equal mass) particles.
The five integrals are all finite. 

We used also the package \texttt{FIRE} \cite{Smi} for an independent check of
the
above reduction. Equation \eqref{Master} shows that the integrals
   $T_1$, $T_2$, and
$T_3$ have to be computed to  order ${\epsilon ^2}$, while it is
 sufficient to compute  $T_4$ and  $T_5$ to  order ${\epsilon}$.

\subsection{Analytic computation of the master integrals}
In the integrands of  $T_1$, $T_2$ and $T_3$           there are no
denominators present, 
 and the  respective  numerator factors
depend just on a subset of phase space momenta. Based on this observation, we
first rewrite the phase space $d \Phi_4$ in terms of the convolution formula 
\begin{equation} \label{MI1_5}
d \Phi_4 ( p_1, p_2, p_3, p_4 ; q ) = \frac{1}{2\pi} \int^{q^2}_{4 m^2}\! dM^2\:
d\Phi_2( p_4, k; q )\, d \Phi_3( p_1, p_2, p_3; k ),
\end{equation}
where $k^2 = M^2$. Using this relation along with standard
identities and integral representations of hypergeometric functions \cite{Pru},
we find that the first three master integrals can be expressed 
in terms of hypergeometric functions ${}_3F_2$:
\begin{eqnarray}
%
%
T_1 & = & \left( q^2 \right)^2  \left( \frac{\mu^2}{q^2} \right)^{3
\epsilon } \bigg\{\, 2^{- 11 + 6 \epsilon }\, \pi ^{-5+3 \epsilon } \,\frac{
\Gamma(1-\epsilon )^4 }{ \Gamma(3-3 \epsilon )\, \Gamma(4-4 \epsilon ) } 
\nonumber \\[0.2cm]
& & \quad {} \times {}_3F_2\!\left( -\frac{1}{2}+\epsilon ,-2+3 \epsilon ,-3+4
\epsilon; \epsilon ,-1+2 \epsilon; z\right)
\nonumber \\[0.2cm]
& & {} + 2^{- 12 + 8 \epsilon }\, \pi ^{- 5 + 3 \epsilon } \, \frac{
\Gamma(1-\epsilon )^3\, \Gamma(-1+\epsilon ) }{ \Gamma( 3 -3 \epsilon )\,
\Gamma(2-2 \epsilon ) } 
\nonumber \\[0.2cm]
& & \quad {} \times z^{1-\epsilon } \, {}_3F_2\!\left( \frac{1}{2},-1+2 \epsilon
,-2+3 \epsilon ; 2-\epsilon ,\epsilon ; z \right)  
\nonumber \\[0.2cm]
& & {} + 2^{- 15 + 10 \epsilon }\, \pi ^{-5+3 \epsilon }\,\frac{
\Gamma(1-\epsilon )\, \Gamma(-1+\epsilon )^2 }{ 
\Gamma(2-2 \epsilon ) }
\nonumber \\[0.2cm]
& & \quad {} \times z^{2-2 \epsilon } \, {}_3F_2\!\left( \frac{3}{2}-\epsilon
,\epsilon,-1+2 \epsilon ; 3-2 \epsilon ,2-\epsilon ; z \right) \bigg\} ,
\\[0.4cm]
%
%
 T_2 & = & \left( q^2 \right)^3 \left( \frac{ \mu^2 }{ q^2 } \right)^{3
\epsilon}  \bigg\{\, 2^{- 12 + 6 \epsilon }\, \pi^{-5 + 3 \epsilon }\, \frac{
\Gamma( 1 - \epsilon )^4 }{ 3\, \Gamma( 3 - 3 \epsilon)\, \Gamma( 4 - 4
\epsilon) }
\nonumber \\[0.2cm]
& & \quad {} \times {}_3F_2 \! \left( - \frac{1}{2} + \epsilon, -3 + 3\epsilon,
-4 + 4 \epsilon ; \epsilon , -2 + 2 \epsilon ;  z \right)
\nonumber \\[0.2cm]
& & {} + 2^{- 13 + 8 \epsilon }\, \pi^{- 5 + 3 \epsilon }\, \frac{
\Gamma( 1 - \epsilon)^3 \Gamma( -1 + \epsilon) }{ 3\, \Gamma( 3 - 3 \epsilon )\,
\Gamma( 2 - 2 \epsilon ) }
\nonumber \\[0.2cm]
& & \quad {} \times z^{ 1 - \epsilon }\, {}_3F_2 \! \left( \frac{1}{2}, -2 + 2
\epsilon, -3 + 3 \epsilon ; 2 - \epsilon , -1 + \epsilon ;  z \right)  
\nonumber \\[0.2cm]
& & {} + 2^{- 16 + 10 \epsilon }\, \pi^{ -5 + 3 \epsilon}\, \frac{
\Gamma( 2 - \epsilon )\, \Gamma( -2 + \epsilon )\,
\Gamma( - 1 + \epsilon ) }{ \Gamma( 2 - 2 \epsilon) } 
\nonumber \\[0.2cm]
& & \quad {} \times z^{ 3 - 2 \epsilon }\, {}_3F_2 \! \left( \frac{5}{2} -
\epsilon, \epsilon, -1 + 2 \epsilon ; 4 - 2 \epsilon , 3 - \epsilon ;  z \right)
\bigg\},
\\[0.4cm]
%
%
T_3 & = &\left( q^2 \right)^3 \left( \frac{\mu^2 }{ q^2
} \right)^{3 \epsilon } \bigg\{  2^{-12 + 6 \epsilon } \pi^{-5+3 \epsilon } \,
\frac{ \Gamma ( 1-\epsilon )^4 }{ \Gamma( 3 - 3 \epsilon
)\, \Gamma ( 4 - 4 \epsilon ) }
\nonumber \\[0.2cm]
& & \quad {} \times {}_3 F_2 \! \left( -\frac{1}{2}+\epsilon ,-2+3 \epsilon
,-4+4 \epsilon ; \epsilon ,-1+2 \epsilon ; z\right)
\nonumber \\[0.2cm]
& & {} + 2^{-11 + 8 \epsilon }\, \pi^{ - 5 + 3 \epsilon }\, \frac{
\Gamma ( 1-\epsilon )^3\, \Gamma (-1+\epsilon ) }{
3 \, \Gamma ( 3-3 \epsilon )\, \Gamma ( 2-2 \epsilon ) }
\nonumber \\[0.2cm]
& & \quad {} \times z^{ 1-\epsilon } \, {}_3 F_2 \! \left( \frac{1}{2},-1+2
\epsilon ,-3+3 \epsilon ; 2-\epsilon ,\epsilon; z \right)
\nonumber \\[0.2cm]
& & {} + 2^{- 15 + 10 \epsilon } \pi^{-5+3 \epsilon }\, \frac{ 
\Gamma ( 1-\epsilon )\, \Gamma( -1+\epsilon )^2  }{ \Gamma ( 2-2 \epsilon ) }
\nonumber \\[0.2cm]
& & \quad {} \times z^{2-2 \epsilon }\, {}_3 F_2 \! \left( 
\frac{3}{2}-\epsilon ,\epsilon ,-2+2 \epsilon ; 3-2 \epsilon
,2-\epsilon ; z \right) \bigg\}.
\end{eqnarray}
We use the package \texttt{HypExp2} \cite{HubMai} for expanding the hypergeometric
functions  in ${\epsilon}$  to the required orders.
For the above three master integrals the result of these expansions is
given in Appendix \ref{appendix master} 
in terms of harmonic polylogarithms
(HPL) \cite{RemVer,Gehrmann:2001pz}. In \cite{Czakon:2011ve} the phase space
volume
$T_1$ was also computed by
 expansion in $\epsilon$ in terms of harmonic polylogarithms.

The remaining two master integrals ${T_4,\,T_5}$ are computed using  the
 differential equations method \cite{Kot,Rem,Gehrmann:1999as}. (For a review see
\cite{ArgMas}.) By differentiating  ${T_4,\,T_5}$ with respect to 
${z=4m^2/q^2}$ we obtain linear combinations of integrals which in turn can be
reduced by IBP identities to the above five master integrals under
consideration. In this way we derive inhomogeneous first order differential
equations in ${z}$ for ${T_4}$ and ${T_5}$. 

The inhomogeneous part of the differential
equation for $T_4$ only depends on the three master integrals
${T_1,\,T_2,\,T_3}$
whose series expansions in ${\epsilon}$ were already derived above in terms of
HPL. We expand this equation for
${T_4}$ in ${\epsilon}$:
\begin{eqnarray}
 T_4 & = & T_4^{(0)}+T_4^{(1)}\epsilon +\mathcal{O} \left( \epsilon ^2 \right).
\end{eqnarray}
In this way we obtain first order differential equations for the coefficients
${T_4^{(0)}}$ and
${T_4^{(1)}}$, whose inhomogeneous parts are determined in terms of HPL. In
each case the general
solution is composed of the general solution of the homogeneous
equation, which contains a constant of
integration, and an integral over the inhomogeneous part.
The differential 
equations for $T_5$ are obtained and solved in a similar way.
 Here the inhomogeneous part depends on the other four master integrals.

The integration over the inhomogeneous parts is carried out using the
package \texttt{HPL} \cite{Mai}. After partial fraction decomposition and
expansion of shuffle products we can write the respective integrand such that all terms
containing HPL are of the form ${k\, f(z) ^j\,H(...,z)}$, where ${k}$ is a
constant, ${ f(z) }$ is one of the functions
${\frac{1}{1-z},\,\frac{1}{z},\,\frac{1}{1+z}}$ appearing to a positive integer
power ${j}$, and where ${H(...,z)}$ is a HPL of weight ${w}$ and argument
${z}$. In the case of ${j=1}$ the primitive is ${k}$ times an HPL of weight
${w+1}$, which follows directly from the definition of the HPL  \cite{RemVer}. In
the remaining cases where ${j\neq1}$ we can perform partial integration and
partial fractioning in sequence until each term is either of the above
form with ${j=1}$ or is just an algebraic function of ${z}$.

In order to fix the constants of integration of the  integral
$T_5$, one can consider the massless limit $z=0$
where $T_5$    remains finite. In this limit
 the integral was computed in \cite{GehGehHei}. We use that  result 
as a boundary condition in order to determine the constants of integration for $T_5$. 

In the case of the differential equations obtained for ${T_4^{(0)}}$ and
${T_4^{(1)}}$, the massless limit $z=0$ and the threshold limit $z=1$ can not be
used as boundary conditions for the determination of the integration
constants. 
 Instead we choose an appropriate  integral which is known to vanish
in the limit $z=1$ and which can be expressed in terms of
the five master integrals by IBP reduction. Via this reduction we use the latter
limit 
as boundary condition for fixing  these integration constants.

For  $T_4$ and $T_5$      the result of their expansion to oder $\varepsilon$
  in terms of harmonic polylogarithms  is also  given in Appendix \ref{appendix master}.

For the integrals $T_4$ and $T_5$  we performed numerical
cross checks using \texttt{VEGAS}
\cite{Lep78}. A strong analytical check of  all five master integrals is
provided by analysis  described in the
subsequent section.

In the following we use the variable
\begin{eqnarray}
y & \equiv & \frac{1-\sqrt{1-z}}{1+\sqrt{1-z}} \, .
\end{eqnarray}
 Inserting our results for the master integrals into
\eqref{Master} we obtain as our main result the 
integrated antenna function expanded in $\epsilon$:
\begin{eqnarray}
%
%
\cB^0_{Q q \bar{q} \bar{Q}} ( q^2, y, \mu, \epsilon ) & = & \left(
\frac{\mu^2}{q^2} \right)^{2 \epsilon} \bigg[ \, \frac{1}{\epsilon^2}
\left\{ -\frac{1}{6}+\left(\frac{1}{6}-\frac{1}{6 (1-y)}-\frac{1}{6
(1+y)}\right) H( 0; y) \right\} 
\nonumber \\[0.2cm]
%
%
& & {} + \frac{1}{ \epsilon }\, \bigg\{  \left(-\frac{8}{9}+\frac{23}{36
(1-y)}-\frac{13}{36 (1+y)}+\frac{1+2 y}{6 \left(1+4 y+y^2\right)}\right)
H( 0; y )
\nonumber \\[0.2cm]
& & -\frac{4}{3}\, H( 1; y) + \left(\frac{4}{3}-\frac{4}{3
(1-y)}-\frac{4}{3 (1+y)}\right) H( 2 ; y )
\nonumber \\[0.2cm]
& & {} + \left(\frac{4}{3}-\frac{4}{3 (1-y)}-\frac{4}{3
(1+y)}\right) H( -1,0 ; y )
\nonumber \\[0.2cm]
& & {} - \left( \frac{1}{3} - \frac{1}{3
(1-y)} - \frac{1}{3 (1+y)}\right) H( 0,0; y )
\nonumber \\[0.2cm]
& & {} - \left( \frac{2}{3} - \frac{2}{3 (1-y)} - \frac{2}{3
(1+y)}\right) H( 1,0; y)
\nonumber \\[0.2cm]
& & {} - \frac{43}{36} - \frac{2 \pi^2 }{9} + \frac{2 \pi ^2}{9
(1-y)}+\frac{2 \pi^2}{9 (1+y)}-\frac{y}{1+4 y+y^2} \bigg\}
\nonumber \\[0.2cm]
%
%
& & - \, \bigg( \frac{ 883 - 15 \pi^2 }{108} - \frac{ 757 - 15 \pi ^2}{108
( 1 - y ) } - \frac{271-15 \pi ^2}{108 (1+y)} - \frac{2+7 y}{3 \left(1+4
y+y^2\right)^2} 
\nonumber \\[0.2cm]
& & {} + \frac{61 + 167 y}{18 \left(1+4 y+y^2\right)} \bigg) H( 0; y)
- \bigg( \frac{14}{9} + \frac{2}{(1-y)^2} -
\frac{95}{36 (1-y)}
\nonumber \\[0.2cm] 
& & { } - \frac{35}{36 (1+y)} - \frac{5+22 y}{6 \left(1+4
y+y^2\right)} \bigg)
H( 0,0; y)
\nonumber \\[0.2cm]
& & {}- \left( \frac{5 \pi ^2}{3} - \frac{5 \pi ^2}{3
(1 - y)} - \frac{5 \pi ^2}{3 (1+y)}\right) H( -1 ; y )
\nonumber \\[0.2cm]
& & {} - \left( \frac{ 86 - 5 \pi^2 }{9} + \frac{5 \pi ^2}{9 (1-y)} + \frac{5
\pi ^2}{9 (1+y)} + \frac{8 y}{1+4 y+y^2}\right) H( 1; y )
\nonumber \\[0.2cm]
& & {} - \left( \frac{64}{9} - \frac{46}{9
(1 - y )} +  \frac{26}{9 (1+y)} - \frac{4 (1+2 y)}{3 \left(1+4
y+y^2\right)}\right) H( 2; y)
\nonumber \\[0.2cm]
& & {} - \left( \frac{8}{3} - \frac{8}{3 (1-y)} - \frac{8}{3
(1+y)}\right) H( 3; y) - \frac{32}{3}\, H( 1,1; y )
\nonumber \\[0.2cm]
& & {} - \left( \frac{20}{3} - \frac{20}{3(1-y)} - \frac{20}{3 (1+y)}\right)
H( -2,0; y )
\nonumber \\[0.2cm]
& & {} - \left( \frac{4}{9} - \frac{28}{9 (1-y)} + \frac{20}{9 (1+y)}\right)
H( -1,0; y )
\nonumber \\[0.2cm]
& & {} + \left(\frac{32}{3} - \frac{32}{3 (1-y)}-\frac{32}{3 (1+y)}\right)
H( -1,2; y)
\nonumber \\[0.2cm]
& & {} - \left(\frac{58}{9}+\frac{13}{18 (1-y)}-\frac{47}{18
(1+y)}-\frac{1+2 y}{3 \left(1+4 y+y^2\right)}\right)
H( 1,0; y)
\nonumber \\[0.2cm]
& & {} - \left(\frac{16}{3} - \frac{16}{3 (1-y)} - \frac{16}{3
(1+y)}\right) H( 1,2; y)
\nonumber \\[0.2cm]
& & {} + \left(\frac{22}{3} - \frac{22}{3
(1 - y)}-\frac{22}{3 (1+y)}\right)
H( 2,0; y )
\nonumber \\[0.2cm]
& & {} +\left(\frac{32}{3} - \frac{32}{3 (1-y)}-\frac{32}{3
(1+y)}\right) 
H( 2,1; y)
\nonumber \\[0.2cm]
& & {} +\left(\frac{28}{3} - \frac{28}{3 (1-y)}-\frac{28}{3 (1+y)}\right)
H( -1,-1,0; y )
\nonumber \\[0.2cm]
& & {} -\left(\frac{4}{3} - \frac{4}{3 (1-y)}-\frac{4}{3
(1+y)}\right)
H( -1,0,0; y)
\nonumber \\[0.2cm]
& & {} - \left(4 - \frac{4}{1 -y}-\frac{4}{1+y}\right)
H( -1,1,0; y)
\nonumber \\[0.2cm]
& & {} +\left(\frac{2}{3} - \frac{2}{3 (1- y)}-\frac{2}{3
(1+y)}\right)
H( 0,0,0 ; y )
\nonumber \\[0.2cm]
& & {} - \left(4 - \frac{4}{1-y} - \frac{4}{1+y}\right)
H( 1,-1,0; y )
\nonumber \\[0.2cm]
& & {} - \left(\frac{2}{3} - \frac{2}{3 (1 - y)} - \frac{2}{3
(1+y)}\right) H( 1,0,0; y )
\nonumber \\[0.2cm]
& & {} - \frac{ 707 - 113 \pi^2 + 468\, \zeta( 3 ) }{108} +\frac{79
\pi ^2+468 \zeta( 3 )}{108 (1+y)} - \frac{77 \pi^2 - 468\, \zeta ( 3 )
}{108
(1 - y)} 
\nonumber \\[0.2cm]
& & {} -\frac{2 (1+4 y)}{\left(1+4 y+y^2\right)^2}+\frac{12-\pi ^2-36 y-2 \pi
^2
y}{6 \left(1+4 y+y^2\right)}+ \cO( \epsilon ) \bigg] .
\label{Antenna function}
\end{eqnarray}

%% file: QQqq_Check.tex
\section{The correction of
  \boldmath $\alpha_s^2 e^2_Q N_f$ to the
ratio R}
\label{check}
As an application and check of our results of Section~3, we consider the
ratio
\begin{equation} \label{Rhad}
 R = \frac{ \sigma(e^+ e^- \to \gamma^\ast \to \, Q \bar{Q} + X )}{
\sigma_\mss{pt} }\,,
\end{equation}
  to order  $\alpha_s^2$ and to lowest order in $\alpha$.
 Here $\sigma_\mss{pt} = e^4 /
(12 \pi q^2 )$ is the massless Born cross section
for $e^+ e^- \to \gamma^\ast \to \mu^+ \mu^-$. 
 In the following, we consider one heavy quark, carrying
the electric charge $e_Q$, and $N_f$
massless  quark flavors. Here
 we are only  interested in the contribution
proportional to $\alpha_s^2 e^2_Q N_f$ 
  to  the  ratio (\ref{Rhad}). This contribution is gauge-invariant
   and IR finite. Apart from the tree-level contributions, which are
   closely related to the integrated antenna function of
   Section~3, this term receives a two-loop contribution which was computed
   within dimensional regularization first in \cite{Bernreuther:2004ih}.
   Using this result and our result of Section~3, we can check
     the IR poles  of  $\cB^0_{Q q \bar{q} \bar{Q}}$.
   Furthermore we compare this contribution to $R$  with the
   previous result
   of  \cite{Hoa}, which was obtained in $d=4$ using  different methods.

Throughout this section we use the subscripts
   $\alpha_s^2 e_Q^2
N_f$ or $\alpha_s^2 N_f$  when referring to the
contribution  of these terms to  a given quantity.
We have
\begin{equation} \label{sigmaR}
 \sigma(e^+ e^- \to \gamma^\ast \to Q \bar{Q} + X )_{ \alpha^2_s e^2_Q N_f } =
\frac{1}{2 q^2}\, \frac{1}{4} \, \frac{e^4 e^2_Q}{ \left( q^2 \right)^2 }\,
L^{\mu\nu} \sum_X H^{Q\bar{Q}X}_{\mu\nu, \, \alpha^2_s N_f } \, ,
\end{equation}
with the lepton tensor
\begin{equation} \label{lepten}
 L_{\mu\nu} = 4 \left( k_{1\mu} k_{2 \nu} + k_{1 \nu} k_{2 \mu} - g_{\mu
\nu}\, k_1 \cdot k_2 \right) \, , 
\end{equation}
where $k_{1\mu}$ and $k_{2\mu}$ denote the momenta of the incoming electron and
positron, $q_\mu = k_{1\mu} +
k_{2\mu}$, and the contributions
\begin{equation} \label{hadten}
 H^X_{\mu \nu,\alpha^2_s N_f} = \left( q_\mu q_\nu - g_{\mu \nu}\, q^2 \right)
\Pi^X_{\alpha^2_s N_f}(q^2,m,\mu,\epsilon)  
\end{equation}
 to the hadron tensor. They will be given below. Performing the tensor contractions in $d = 4 - 2 \epsilon$ 
 dimensions, one obtains
\begin{equation} \label{RPi}
R_{ \alpha_s^2 e^2_Q N_f } = 6 \pi e^2_Q \left( 1 - \epsilon \right)
\sum_X \Pi^{Q\bar{Q}X}_{\alpha_s^2 N_f} .
\end{equation}
The contribution to (\ref{RPi}) from the $Q \bar{Q} q \bar{q}$
final state is closely related to the integrated antenna function $\cB^0_{Q q
\bar{q} \bar{Q}}$ given in eq. \eqref{Antenna function}. After restoring all
couplings and color factors we find 
\begin{equation} \label{PiQQqq}
\Pi^{Q\bar{Q}q\bar{q}}_{\alpha_s^2 N_f} 
= \frac{ ( 4 \pi
\alpha_s )^2 4 C_F N_c T_R N_f}{ q^2 ( 3 - 2 \epsilon ) }\, \frac{ P_2 ( q^2 , m
)\, | \cM_{Q \bar{Q} } |^2 }{ \left( 8 \pi^2 ( 4 \pi )^{- \epsilon} e^{ \gamma_E
\epsilon } \right)^2 }\, \cB^0_{Q q \bar{q} \bar{Q} } ( q^2, y, \mu, \epsilon )
\,,
\end{equation}
where $C_F = (N_c^2 - 1)/( 2 N_c)$ and $T_R = \einhalb$. The 
expressions for $| \cM_{Q \bar{Q} }|^2$ and $P_2( q^2, m)$ 
are given in  \eqref{MQQ} and \eqref{PS2}, respectively. The second
normalization factor in \eqref{PiQQqq} is obtained in straightforward fashion; it reflects the
 relation between the decay rate of a virtual photon and the integrated antenna
function, see  \eqref{ma}, \eqref{gm_19} and \eqref{gm_18}. In our
calculation of  the antenna function the hadronic tensor \eqref{hadten} was
contracted with $-g_{\mu \nu}$  instead of $L_{\mu\nu}$. This is
corrected by  the additional factor $1/(q^2 (3 - 2\epsilon))$.

The two-particle contribution $\Pi^{Q \bar{Q}}$ can be expressed by  the
Dirac and Pauli heavy quark form  form factors $F_1$ and $F_2$.
\begin{eqnarray}
 \Pi^{Q \bar{Q}}_{\alpha_s^2 N_f} 
& = & \frac{N_c P_2 ( q^2, m ) }{ 3 - 2 \epsilon } \bigg[  4  \left(1+\frac{2
y}{(1+y)^2}-\epsilon \right)  | F_1 |^2_{\alpha_s^2 N_f} 
\nonumber \\[0.2cm]
& & { } + \frac{ \left(1+y^2+y (10-8 \epsilon
)\right) }{2 y} \, | F_2 |^2_{\alpha_s^2 N_f}  + 4 ( 3 - 2 \epsilon ) \text{Re}(
F_1 F_2^\ast )_{\alpha_s^2 N_f} \bigg] \, . \label{eq:conQQ2l}
\end{eqnarray}
 The contributions required in  (\ref{eq:conQQ2l})
 can  be read off from the expressions  for the UV-renormalized
form factors above threshold given in \cite{Bernreuther:2004ih}.

In \cite{Bernreuther:2004ih} the renormalization constants for the heavy quark
mass and wave function were defined in the on-shell scheme, whereas the
renormalization of the strong coupling constant and the gluon wave-function
was performed in the $\overline{\text{MS}}$ scheme. In order to obtain
an LSZ residue equal to one, we apply
on-shell renormalization for the external gluon, too. (Nominally, this
  avoids contributions from three-particle cuts.)
This change of the renormalization scheme as compared to
\cite{Bernreuther:2004ih} 
leaves the two-loop contributions  $\propto \alpha_s^2 N_f$ 
 unchanged.  However, it changes the 
  $QQg$ renormalization
           constant $Z_{1F} ( \epsilon,\mu^2/m^2)$ as compared to the
           one of  \cite{Bernreuther:2004ih} 
 at the one-loop level by the additional term 
\begin{equation}
\delta Z_{1F, \alpha_s N_f}
= \frac{\alpha_s N_f T_R
\left( 4\pi \right)^\epsilon}{6\pi \epsilon} \Gamma \left( 1+\epsilon
\right) \, .
\end{equation}
This change induces a  counterterm contribution proportional to
        $\alpha_s^2 N_f$  from the three particle $Q \bar{Q} g$ final
state, which reads:
\begin{equation} \label{eq:3pcountc}
 \Pi^{Q\bar{Q}g}_{\alpha^2_S N_f}
= \frac{ (4 \pi
\alpha_s) \mu^{2\epsilon} 2 C_F N_c }{ q^2 ( 3 - 2 \epsilon ) } \, P_2 ( q^2 , m
)\, | \cM_{Q \bar{Q} } |^2  \left( 2 \delta Z_{1F, \alpha_s N_f}\, \cA^0_{Q g
\bar{Q} } \right)
\,,
\end{equation}
where $\cA^0_{Q g \bar{Q}}$ is the integrated massive tree-level three parton
quark-antiquark antenna as given, e.g.,  in
\cite{GehrmannDeRidder:2009fz}.

Adding the contributions  (\ref{PiQQqq}), (\ref{eq:conQQ2l}), and  (\ref{eq:3pcountc})
 to (\ref{RPi}), all IR poles cancel. This provides a strong  check 
  for the IR divergent part of the integrated antenna function given in
\eqref{Antenna function}. \\
Next we compare our result for $R_{ \alpha_s^2 e^2_Q N_f}$ with the
result of  
   of  \cite{Hoa}, which was obtained in $d=4$ using  different techniques.
Introducing the QCD coupling $\alpha_s$ and the appropriate color factor
 in  eq.\:(42) of ref.\:\cite{Hoa} the relevant part of this
  equation becomes
\begin{eqnarray}
 R_{ \alpha_s^2 e^2_Q N_f } & = & \left( \frac{ \alpha_s }{ \pi } \right)^2 e_Q^2 C_F T_R N_c N_f 
\nonumber \\[0.2cm]
& & {} \times \left[ - \frac{1}{3}\, W
\, \ln\left( \frac{ \mu^2 }{ q^2 } \right) + f^{(0)}_R + w ( 3 - w^2 )  \left(
f^{(0)}_1 + f^{(0)}_2 \right) + w^3 f^{(0)}_2 \right]\!,
\label{RHoang}
\end{eqnarray}
where
\begin{equation}
 w = \frac{ 1 - y }{1 + y } = \sqrt{1-z}\, , 
\end{equation}
 and the explicit expressions of the functions $W$, $f^{(0)}_R$, $f^{(0)}_1$,
and $f^{(0)}_2$ can be found in reference \cite{Hoa}. 
    In \cite{Hoa} the
result for $f^{(0)}_1$ is expressed  in  terms of the integrals
\begin{eqnarray}
T_2(\eta,\,\xi)& = &\int _0 ^1
\! dx\: \frac{\textrm{arctan}(\xi\,x)}{x^2+\eta^2},
\nonumber \\
T_2^ \star (\eta,\,\xi)& = &\int _0 ^1 \! dx\: \frac{\textrm{ln}(x^2+\xi
^2)}{x^2+\eta^2}, 
\\
T_3(\eta,\,\xi,\,\chi)& = &\int _0 ^1 \! dx\: \frac{\textrm{ln}(x^2+\xi ^2)\, 
\textrm{arctan}(\chi x)}{x^2+\eta^2}.\nonumber 
\end{eqnarray}
In order to be able to compare our result with (\ref{RHoang}) in
analytic fashion, we have computed 
  the particular combination of these integrals, which appears 
 in the function $f^{(0)}_1$, 
in terms of polylogarithms. We find 
\begin{eqnarray}
\lefteqn{ -T_3 \left( 1,\,0,\,w \right) + T_3 \left( 1,\,\frac{1}{w},\,w \right)
- T_3 \left( 1,w,\,\frac{1}{w} \right) + 2\, \textrm{ln} (w)\, T_2
\left(1,\,w \right) -\frac{\pi}{2}\, T_2 ^\ast
\left( 1,\,\frac{1}{w}\right) } \quad \nonumber \\[0.2cm]
& = & \text{Li}_3\! \left( -\frac{w}{1-w} \right) - \text{Li}_3\! \left(
\frac{w}{1+w} \right) + \ln( w ) \left( \text{Li}_2\! \left( \frac{w}{1+w}
\right) - \text{Li}_2\! \left( - \frac{w}{1  - w}
\right) \right) 
\nonumber \\[0.2cm]
& & {} + \frac{1}{6}\, \ln^3 (1 + w) - \frac{1}{6}\, \ln^3(1-w) - \frac{ \pi^2
 }{3} \, \ln(1+w) - \frac{ \pi^2 }{6}\, \ln(1-w) 
\nonumber \\[0.2cm]
& & {} + \frac{ \pi^2 }{4}\, \ln(w) + \frac{1}{2}\, \ln^2(w) \left( \ln(1-w) -
\ln(1+w) \right) + \pi G \, ,  
\label{Tanalytic}
\end{eqnarray}
where $G$ is Catalan's constant. With this formula we find
agreement\footnote{In the course
 of this comparison, we found that  eq. (23) of \cite{Hoa}
  contains a typographical error.  In the fifth line of this
equation, $\textrm{ln}\,p^2$ must be replaced by  $\textrm{ln}^2\,p$.}
  between our result  (\ref{RPi})  and  the result (\ref{RHoang}) of  \cite{Hoa}.
%
%

%% file: QQqq_conclusion_acknowledgements.tex
%
%
%
\section{Summary and Outlook}
As a first step towards extending the antenna subtraction method
 to NNLO QCD reactions with massive quarks, we 
 have determined the real radiation  antenna function and its
 integrated counterpart for reactions of the type $S \to Q{\bar Q} q
 {\bar q}$, where $S$ denotes an uncolored initial state.
  We were able to determine the integrated antenna function in
  completely analytic fashion, namely in terms of harmonic
  polylogarithms, for which efficient evaluation codes are available.
 We checked our results by computing  the contribution
proportional to $\alpha_s^2 e^2_Q N_f$ to the inclusive heavy-quark
pair production cross section in $e^+e^-$ annihilation via a virtual photon and
 by comparison with results in the literature. 

An obvious next step in this line of investigation is the determination
of the antenna function and its integrated version for $S \to Q{\bar Q} g g$. 
The results of this paper indicate that for the  $Q{\bar Q} g g$ final state,
the integrated antenna function can also be obtained analytically in a
relatively compact form. 
%
%
\acknowledgments 
We are indebted to Thomas Gehrmann and Aude Gehrmann-De Ridder 
  for helpful discussions and comments on the manuscript. We wish to
thank Karl Waninger for a comparison of matrix elements, Tobias Huber
and Daniel Maitre for an E-mail exchange on \cite{HubMai}, and Andre Hoang and
Thomas Teubner for an E-mail exchange on \cite{Hoa}. This work was  supported 
by Deutsche Forschungsgemeinschaft (DFG), SFB/TR9 and by BMBF. The figures were
generated using Jaxodraw \cite{Binosi:2003yf}, based on Axodraw
\cite{Vermaseren:1994je}.

%% file: QQqq_Appendix_Masters.tex
%
%
%
%
\section{The master integrals} 
\label{appendix master}
In this appendix we give analytic results for the five master integrals 
${T_1,\,T_2,\,T_3,\,T_4,\,T_5}$ to the required orders in ${\epsilon}$. The
integrals are given in terms of the variable
\begin{equation}
 y \equiv \frac{ 1 - \sqrt{ 1 - \frac{4 m^2 }{ q^2 } } }{  1 + \sqrt{ 1 -
\frac{4 m^2 }{ q^2 } } }\,.
\end{equation}
The harmonic polylogarithms are given in the notation of
\cite{RemVer}. 
The $\epsilon$-expansion is needed to order $\epsilon^2$ for $T_1$, $T_2$ and
$T_3$ and to order $\epsilon$ for $T_4$ and $T_5$. For convenience, we define
\begin{equation}
T_1( q^2, y ,\mu^2, \epsilon ) = C^3( \epsilon ) \left( \frac{ \mu^2 }{
q^2 } \right)^{3 \epsilon} \frac{ (q^2)^2 }{2^{11} \pi^5 (1 +
y)^6 } \, \cT_1 ( y, \epsilon ),
\end{equation}
with
\begin{equation}
 C( \epsilon ) = \left( \frac{ 4 \pi }{ e^{ \gamma_E } } \right)^\epsilon\,.
\end{equation}
We find
\begin{eqnarray}
 \cT_1 ( y, \epsilon ) & = &
%
%
-\frac{1}{12} (1+y) \left(-1-23 y+\left(-34+4 \pi
^2\right) y^2+\left(34+4 \pi ^2\right) y^3+23 y^4+y^5\right) \nonumber \\
& & {} +y \left(1+5 y+6
y^2+5 y^3+y^4\right) \, H(0 ;y)-4 y^2 (1+y)^2 \, H(-1,0 ;y) \nonumber \\
& &{} +2y^2 (1+y)^2 \, H(0,0 ;y) \big\}
\nonumber \\
%
%
& & { } + \epsilon \bigg\{ 4 y^2 (1+y)^2 \, H(-3 ;y)+2 y \left(1+5 y+6
y^2+5 y^3+y^4\right) \, H(-2 ;y)\nonumber \\
& & {} -\frac{1}{6} (1+y) \left(-1-23 y-2
\left(17+8 \pi ^2\right) y^2+\left(34-16 \pi ^2\right) y^3+23 y^4  +y^5\right)
\, H(-1 ;y) \nonumber \\
& & {} -\frac{1}{6} y \left(-45+\left(-183+8 \pi ^2\right) y+2
\left(-35+8 \pi ^2\right) y^2+\left(45+8 \pi ^2\right) y^3 \right. \nonumber \\
& & {} \left. +51 y^4+4 y^5\right)
\, H(0 ;y)-\frac{5}{6} \left(-1-24 y-57 y^2+57 y^4+24 y^5+y^6\right)
\, H(1 ;y) \nonumber \\
& & {} +10 y \left(1+5 y+6 y^2+5 y^3+y^4\right)
\, H(2 ;y)+20 y^2 (1+y)^2 \, H(3 ;y) \nonumber \\
& & {} +28 y^2 (1+y)^2
\, H(-2,0 ;y)-8 y^2 (1+y)^2 \, H(-1,-2 ;y) \nonumber \\
& & {} +2 y \left(5+17 y+16
y^2+17 y^3+5 y^4\right) \, H(-1,0 ;y)-40 y^2 (1+y)^2
\, H(-1,2 ;y) \nonumber \\
& & {} -y (1+y)^2 \left(1-5 y+y^2\right) \, H(0,0 ;y)-56
y^2 (1+y)^2 \, H(-1,-1,0 ;y) \nonumber \\
& & {} +12 y^2 (1+y)^2 \, H(-1,0,0 ;y)-6
y^2 (1+y)^2 \, H(0,0,0 ;y) \nonumber \\
& & {} +\frac{1}{72} \left(71-24 \left(-71+3 \pi
^2\right) y-24 \left(71+3 \pi ^2\right) y^5-71 y^6 \right. \nonumber \\
& & {}  +y^2 \left(4047-456 \pi ^2-720
\, \zeta(3)\right)-120 y^3 \left(5 \pi ^2+12 \, \zeta(3)\right) \nonumber \\
& & {}  \left. -3 y^4
\left(1349+152 \pi ^2+240 \, \zeta(3)\right)\right) \bigg\}
\nonumber \\
%
%
& & { } + \epsilon^2 \bigg\{ {} - \frac{7}{432} \left(-445+9 \pi
^2\right)+\frac{5}{432} \left(-623+75 \pi ^2\right) y^6+\left(-12 y^2-24 y^3 \right. \nonumber \\
& & {} \left. -12
y^4\right) \, H(-4 ;y)+\left(-2 y+6 y^2+16 y^3+6 y^4-2 y^5\right)
\, H(-3 ;y) \nonumber \\
& & {} +\left(15 y-\frac{1}{3} \left(-183+80 \pi ^2\right)
y^2-\frac{10}{3} \left(-7+16 \pi ^2\right) y^3-\frac{5}{3} \left(9+16 \pi
^2\right) y^4 \right. \nonumber \\
& & {} \left. -17 y^5-\frac{4 y^6}{3}\right)
\, H(-2 ;y) \nonumber \\
& & {} +\left(\frac{355}{36}+\frac{710 y}{3}+\frac{6745
y^2}{12}-\frac{6745 y^4}{12}-\frac{710 y^5}{3}-\frac{355 y^6}{36}\right)
\, H(1 ;y) \nonumber \\
& & {} +\left(75 y+305 y^2+\frac{350 y^3}{3}-75 y^4-85 y^5-\frac{20
y^6}{3}\right) \, H(2 ;y) \nonumber \\
& & {} +\left(-10 y+30 y^2+80 y^3+30 y^4-10
y^5\right) \, H(3 ;y) \nonumber \\
& & {} +\left(-60 y^2-120 y^3-60 y^4\right)
\, H(4 ;y) \nonumber \\
& & {} +\left(8 y^2+16 y^3+8 y^4\right)
\, H(-3,-1 ;y) \nonumber \\
& & {} +\left(-132 y^2-264 y^3-132 y^4\right)
\, H(-3,0 ;y) \nonumber \\
& & {} +\left(40 y^2+80 y^3+40 y^4\right)
\, H(-3,1 ;y)\nonumber \\
& & {} +\left(56 y^2+112 y^3+56 y^4\right)
\, H(-2,-2 ;y)\nonumber \\
& & {} +\left(4 y+20 y^2+24 y^3+20 y^4+4 y^5\right)
\, H(-2,-1 ;y)\nonumber \\
& & {} +\left(-38 y-78 y^2-32 y^3-78 y^4-38 y^5\right)
\, H(-2,0 ;y)\nonumber \\
& & {} +\left(20 y+100 y^2+120 y^3+100 y^4+20 y^5\right)
\, H(-2,1 ;y)\nonumber \\
& & {} +\left(280 y^2+560 y^3+280 y^4\right)
\, H(-2,2 ;y)\nonumber \\
& & {} +\left(24 y^2+48 y^3+24 y^4\right)
\, H(-1,-3 ;y)\nonumber \\
& & {} +\left(20 y+68 y^2+64 y^3+68 y^4+20 y^5\right)
\, H(-1,-2 ;y)\nonumber \\
& & {} +\left(\frac{1}{3}+8 y+\frac{1}{3} \left(57+160 \pi
^2\right) y^2+\frac{320 \pi ^2 y^3}{3}+\frac{1}{3} \left(-57+160 \pi ^2\right)
y^4 \right. \nonumber \\
& & {} \left.-8 y^5-\frac{y^6}{3}\right) \, H(-1,-1 ;y) \nonumber \\
& & {} +\left(-\frac{7}{3}+19
y-\frac{1}{3} \left(-408+7 \pi ^2\right) y^2-\frac{2}{3} \left(-136+7 \pi
^2\right) y^3-\frac{1}{3} \left(-180+7 \pi ^2\right) y^4 \right. \nonumber \\
& & {} \left.-13 y^5-\frac{11
y^6}{3}\right) \, H(-1,0 ;y) \nonumber \\
& & {} +\left(\frac{5}{3}+40 y+95 y^2-95 y^4-40
y^5-\frac{5 y^6}{3}\right) \, H(-1,1 ;y) \nonumber \\
& & {} +\left(100 y+340 y^2+320
y^3+340 y^4+100 y^5\right) \, H(-1,2 ;y) \nonumber \\
& & {} +\left(120 y^2+240 y^3+120
y^4\right) \, H(-1,3 ;y) \nonumber \\
& & {} +\left(-\frac{15 y}{2}+\frac{1}{6} \left(-75+7
\pi ^2\right) y^2+\frac{1}{3} \left(4+7 \pi ^2\right) y^3+\frac{1}{6} \left(39+7
\pi ^2\right) y^4 \right. \nonumber \\
& & {} \left. +\frac{y^5}{2}+\frac{y^6}{3}\right)
\, H(0,0 ;y) \nonumber \\
& & {} +\left(\frac{5}{3}+40 y+95 y^2-95 y^4-40 y^5-\frac{5
y^6}{3}\right) \, H(1,-1 ;y) \nonumber \\
& & {} +\left(\frac{10}{3}+80 y+190 y^2-190 y^4-80
y^5-\frac{10 y^6}{3}\right) \, H(1,0 ;y) \nonumber \\
& & {} +\left(\frac{25}{3}+200 y+475
y^2-475 y^4-200 y^5-\frac{25 y^6}{3}\right) \, H(1,1 ;y) \nonumber \\
& & {} +\left(20 y+100
y^2+120 y^3+100 y^4+20 y^5\right) \, H(2,-1 ;y) \nonumber \\
& & {} +\left(40 y+200 y^2+240
y^3+200 y^4+40 y^5\right) \, H(2,0 ;y) \nonumber \\
& & {} +\left(100 y+500 y^2+600 y^3+500
y^4+100 y^5\right) \, H(2,1 ;y) \nonumber \\
& & {} +\left(40 y^2+80 y^3+40 y^4\right)
\, H(3,-1 ;y)\nonumber \\
& & {} +\left(80 y^2+160 y^3+80 y^4\right)
\, H(3,0 ;y)+\left(200 y^2+400 y^3+200 y^4\right)
\, H(3,1 ;y)\nonumber \\
& & {} +\left(296 y^2+592 y^3+296 y^4\right)
\, H(-2,-1,0 ;y)\nonumber \\
& & {} +\left(-36 y^2-72 y^3-36 y^4\right)
\, H(-2,0,0 ;y)\nonumber \\
& & {} +\left(-16 y^2-32 y^3-16 y^4\right)
\, H(-1,-2,-1 ;y)\nonumber \\
& & {} +\left(264 y^2+528 y^3+264 y^4\right)
\, H(-1,-2,0 ;y)\nonumber \\
& & {} +\left(-80 y^2-160 y^3-80 y^4\right)
\, H(-1,-2,1 ;y)\nonumber \\
& & {} +\left(-112 y^2-224 y^3-112 y^4\right)
\, H(-1,-1,-2 ;y)\nonumber \\
& & {} +\left(92 y+236 y^2+160 y^3+236 y^4+92 y^5\right)
\, H(-1,-1,0 ;y)\nonumber \\
& & {} +\left(-560 y^2-1120 y^3-560 y^4\right)
\, H(-1,-1,2 ;y)\nonumber \\
& & {} +\left(-6 y+18 y^2+48 y^3+18 y^4-6 y^5\right)
\, H(-1,0,0 ;y)\nonumber \\
& & {} +\left(-80 y^2-160 y^3-80 y^4\right)
\, H(-1,2,-1 ;y)\nonumber \\
& & {} +\left(-160 y^2-320 y^3-160 y^4\right)
\, H(-1,2,0 ;y)\nonumber \\
& & {} +\left(-400 y^2-800 y^3-400 y^4\right)
\, H(-1,2,1 ;y)\nonumber \\
& & {} +\left(y-19 y^2-36 y^3-19 y^4+y^5\right)
\, H(0,0,0 ;y)\nonumber \\
& & {} +\left(-592 y^2-1184 y^3-592 y^4\right)
\, H(-1,-1,-1,0 ;y)\nonumber \\
& & {} +\left(72 y^2+144 y^3+72 y^4\right)
\, H(-1,-1,0,0 ;y)\nonumber \\
& & {} +\left(-28 y^2-56 y^3-28 y^4\right)
\, H(-1,0,0,0 ;y)\nonumber \\
& & {} +\left(14 y^2+28 y^3+14 y^4\right)
\, H(0,0,0,0 ;y)\nonumber \\
& & {} +\left(\frac{71}{36}-\frac{2}{3}
\left(-71+16 \pi ^2\right) y-\frac{2}{3} \left(71+16 \pi ^2\right) y^5-\frac{71
y^6}{36} \right. \nonumber \\
& & {}  -\frac{1}{12} y^2 \left(-1349+512 \pi ^2-384
\, \zeta(3)\right) \nonumber \\
& & {} \left.-\frac{1}{12} y^4 \left(1349+512 \pi ^2-384
\, \zeta(3)\right)-\frac{8}{3} y^3 \left(17 \pi ^2-24
\, \zeta(3)\right)\right) \, H(-1 ;y) \nonumber \\
& & {} + \left(-\frac{3}{4} \left(-47+\pi
^2\right) y 
-\frac{1}{12} \left(1849+9 \pi ^2\right) y^5-\frac{71
y^6}{9}-\frac{1}{12} y^2 \left(-1121+109 \pi ^2+192
\, \zeta(3)\right) \right. \nonumber \\
& & {} \left.-\frac{1}{12} y^4 \left(4275+109 \pi ^2+192
\, \zeta(3)\right)-\frac{1}{18} y^3 \left(3293+249 \pi ^2+576
\, \zeta(3)\right)\right) \, H(0 ;y) \nonumber \\
& & {} +\frac{1}{144} y^2 \left(59185-6021 \pi ^2+76 \pi
^4-18720 \, \zeta(3)\right)  \nonumber \\
& & {} +\frac{1}{144} y^4 \left(-59185+2301 \pi ^2+76 \pi
^4-18720 \, \zeta(3)\right) \nonumber \\
& & {} +\frac{1}{18} y^3 \left(-249 \pi ^2+19 \pi ^4-3204
\, \zeta(3)\right)\nonumber \\
& & {} +\frac{1}{18} y^5 \left(-3115+240 \pi ^2-324
\, \zeta(3)\right) \nonumber \\
& & {} -\frac{1}{18} y \left(-3115+198 \pi ^2+324
\, \zeta(3)\right) \bigg\}.
\end{eqnarray}
\begin{equation}
T_2( q^2, y ,\mu, \epsilon ) = C^3( \epsilon ) \left( \frac{ \mu^2 }{
q^2 } \right)^{3 \epsilon} \frac{ (q^2)^3 }{ 2^{11} \pi^5 (1 +
y )^8 } \, \cT_2 ( y, \epsilon )\,,
\end{equation}
with
\begin{eqnarray}
 \cT_2 ( y , \epsilon ) & = & 
%
%
-\frac{1}{72} (1+y) \left(-1-23 y+67 y^2+\left(149-24 \pi ^2\right)
y^3-\left(149+24 \pi ^2\right) y^4-67 y^5 \right. \nonumber \\
& & {} \left. +23 y^6+y^7\right)+\frac{1}{6}
\left(y-13 y^3-14 y^4-13 y^5+y^7\right) \, H(0 ;y) \nonumber \\
& & {} +4 y^3 (1+y)^2
\, H(-1,0 ;y)-2 y^3 (1+y)^2 \, H(0,0 ;y)
\nonumber \\
%
%
& & {} + \epsilon \bigg\{ -4 y^3 (1+y)^2 \, H(-3 ;y)+\frac{1}{3}
\left(y-13 y^3-14 y^4-13 y^5+y^7\right) \, H(-2 ;y) \nonumber \\
& & {} -\frac{1}{36} (1+y)
\left(-1-23 y+67 y^2+\left(149+96 \pi ^2\right) y^3+\left(-149+96 \pi ^2\right)
y^4-67 y^5  \right. \nonumber \\
& & \left. +23 y^6+y^7\right) \, H(-1 ;y) \nonumber \\
& & {} +\frac{1}{36} y \left(45-30
y+\left(-545+48 \pi ^2\right) y^2+\left(-6+96 \pi ^2\right) y^3+\left(319+48 \pi
^2\right) y^4  \right. \nonumber \\
& & {} \left. +146 y^5-51 y^6-4 y^7\right) \, H(0 ;y) \nonumber \\
& & {} -\frac{5}{36}
\left(-1-24 y+44 y^2+216 y^3-216 y^5-44 y^6+24 y^7+y^8\right)
\, H(1 ;y) \nonumber \\
& & {} +\frac{5}{3} \left(y-13 y^3-14 y^4-13 y^5+y^7\right)
\, H(2 ;y)-20 y^3 (1+y)^2 \, H(3 ;y) \nonumber \\
& & {} -28 y^3 (1+y)^2
\, H(-2,0 ;y)+8 y^3 (1+y)^2 \, H(-1,-2 ;y) \nonumber \\
& & {} +\frac{1}{3} y
\left(5-23 y^2+4 y^3-23 y^4+5 y^6\right) \, H(-1,0 ;y)+40 y^3 (1+y)^2
\, H(-1,2 ;y) \nonumber \\
& & {} -\frac{1}{6} y (1+y)^2 \left(1-2 y+32 y^2-2 y^3+y^4\right)
\, H(0,0 ;y)+56 y^3 (1+y)^2 \, H(-1,-1,0 ;y) \nonumber \\
& & {} -12 y^3 (1+y)^2
\, H(-1,0,0 ;y)+6 y^3 (1+y)^2 \, H(0,0,0 ;y) \nonumber \\
& & {} +\frac{1}{432}
\left(71-24 \left(-71+3 \pi ^2\right) y-2704 y^2+2704 y^6-24 \left(71+3 \pi
^2\right) y^7-71 y^8 \right. \nonumber \\
& & {} +24 y^3 \left(-601+60 \pi ^2+180 \, \zeta(3)\right)+24
y^5 \left(601+60 \pi ^2+180 \, \zeta(3)\right) \nonumber \\
& & {}\left. +24 y^4 \left(79 \pi ^2+360
\, \zeta(3)\right)\right) \bigg\}
\nonumber \\
%
%
& & {} + \epsilon^2 \bigg\{ 12 y^3 (1+y)^2 \, H(-4 ;y)-\frac{1}{3} y
(1+y)^2 \left(1-2 y+32 y^2-2 y^3+y^4\right) \, H(-3 ;y) \nonumber \\
& & {} +\frac{1}{18} y
\left(45-30 y+5 \left(-109+96 \pi ^2\right) y^2+\left(-6+960 \pi ^2\right)
y^3+\left(319+480 \pi ^2\right) y^4 \right. \nonumber \\
& & {} \left. +146 y^5-51 y^6-4 y^7\right)
\, H(-2 ;y) \nonumber \\
& & {} -\frac{5}{216} \left(-71-1704 y+2704 y^2+14424 y^3-14424
y^5-2704 y^6+1704 y^7+71 y^8\right) \, H(1 ;y)\nonumber \\
& & {} -\frac{5}{18} y
\left(-45+30 y+545 y^2+6 y^3-319 y^4-146 y^5+51 y^6+4 y^7\right)
\, H(2 ;y)\nonumber \\
& & {} -\frac{5}{3} \left(y+29 y^3+60 y^4+29 y^5+y^7\right)
\, H(3 ;y)+60 y^3 (1+y)^2 \, H(4 ;y)\nonumber \\
& & {} -8 y^3 (1+y)^2
\, H(-3,-1 ;y)+132 y^3 (1+y)^2 \, H(-3,0 ;y)\nonumber \\
& & {} -40 y^3 (1+y)^2
\, H(-3,1 ;y)-56 y^3 (1+y)^2 \, H(-2,-2 ;y)\nonumber \\
& & {} +\frac{2}{3}
\left(y-13 y^3-14 y^4-13 y^5+y^7\right) \, H(-2,-1 ;y)\nonumber \\
& & {} -\frac{1}{3} y
\left(19+47 y^2+252 y^3+47 y^4+19 y^6\right) \, H(-2,0 ;y)\nonumber \\
& & {} +\frac{10}{3}
\left(y-13 y^3-14 y^4-13 y^5+y^7\right) \, H(-2,1 ;y)\nonumber \\
& & {} -280 y^3 (1+y)^2
\, H(-2,2 ;y)-24 y^3 (1+y)^2 \, H(-1,-3 ;y)\nonumber \\
& & {} +\frac{2}{3} y
\left(5-23 y^2+4 y^3-23 y^4+5 y^6\right) \, H(-1,-2 ;y)\nonumber \\
& & {} -\frac{1}{18}
(1+y) \left(-1-23 y+67 y^2+\left(149+960 \pi ^2\right) y^3+\left(-149+960 \pi
^2\right) y^4 \right. \nonumber \\
& & {} \left. -67 y^5+23 y^6+y^7\right) \, H(-1,-1 ;y)\nonumber \\
& & {} +\frac{1}{18}
\left(-7+57 y+182 y^2+7 \left(-85+6 \pi ^2\right) y^3+6 \left(87+14 \pi
^2\right) y^4 \right. \nonumber \\
& & {} \left. +\left(269+42 \pi ^2\right) y^5+358 y^6-39 y^7-11 y^8\right)
\, H(-1,0 ;y) \nonumber \\
& & {} -\frac{5}{18} \left(-1-24 y+44 y^2+216 y^3-216 y^5-44
y^6+24 y^7+y^8\right) \, H(-1,1 ;y) \nonumber \\
& & {} +\frac{10}{3} y \left(5-23 y^2+4
y^3-23 y^4+5 y^6\right) \, H(-1,2 ;y)-120 y^3 (1+y)^2
\, H(-1,3 ;y) \nonumber \\
& & {} +\frac{1}{36} y (1+y) \left(-45+51 y-2 \left(62+21 \pi
^2\right) y^2-2 \left(211+21 \pi ^2\right) y^3 \right. \nonumber \\
& & {} \left. -83 y^4+y^5+2 y^6\right)
\, H(0,0 ;y) \nonumber \\ 
& & {} -\frac{5}{18} \left(-1-24 y+44 y^2+216 y^3-216 y^5-44
y^6+24 y^7+y^8\right) \, H(1,-1 ;y) \nonumber \\
& & {} -\frac{5}{9} \left(-1-24 y+44
y^2+216 y^3-216 y^5-44 y^6+24 y^7+y^8\right) \, H(1,0 ;y) \nonumber \\
& & {} -\frac{25}{18}
\left(-1-24 y+44 y^2+216 y^3-216 y^5-44 y^6+24 y^7+y^8\right)
\, H(1,1 ;y) \nonumber \\
& & {} +\frac{10}{3} \left(y-13 y^3-14 y^4-13 y^5+y^7\right)
\, H(2,-1 ;y) \nonumber \\
& & {} +\frac{20}{3} \left(y-13 y^3-14 y^4-13 y^5+y^7\right)
\, H(2,0 ;y) \nonumber \\
& & {} +\frac{50}{3} \left(y-13 y^3-14 y^4-13 y^5+y^7\right)
\, H(2,1 ;y)-40 y^3 (1+y)^2 \, H(3,-1 ;y) \nonumber \\
& & {} -80 y^3 (1+y)^2
\, H(3,0 ;y)-200 y^3 (1+y)^2 \, H(3,1 ;y) \nonumber \\
& & {} -296 y^3 (1+y)^2
\, H(-2,-1,0 ;y)+36 y^3 (1+y)^2 \, H(-2,0,0 ;y) \nonumber \\
& & {} +16 y^3 (1+y)^2
\, H(-1,-2,-1 ;y)-264 y^3 (1+y)^2 \, H(-1,-2,0 ;y) \nonumber \\
& & {} +80 y^3
(1+y)^2 \, H(-1,-2,1 ;y)+112 y^3 (1+y)^2
\, H(-1,-1,-2 ;y) \nonumber \\
& & {} +\frac{2}{3} y \left(23-5 y^2+196 y^3-5 y^4+23
y^6\right) \, H(-1,-1,0 ;y)+560 y^3 (1+y)^2 \, H(-1,-1,2 ;y) \nonumber \\
& & {} -y
(1+y)^2 \left(1-2 y+32 y^2-2 y^3+y^4\right) \, H(-1,0,0 ;y)  +80 y^3
(1+y)^2 \, H(-1,2,-1 ;y)\nonumber \\
& & {}+160 y^3 (1+y)^2  \, H(-1,2,0 ;y) \nonumber \\
& & {} +400
y^3 (1+y)^2 \, H(-1,2,1 ;y)+\frac{1}{6} \left(y+113 y^3+208 y^4+113
y^5+y^7\right) \, H(0,0,0 ;y)\nonumber \\
& & {}+592 y^3 (1+y)^2
\, H(-1,-1,-1,0 ;y)-72 y^3 (1+y)^2 \, H(-1,-1,0,0 ;y)\nonumber \\
& & {}+28 y^3
(1+y)^2 \, H(-1,0,0,0 ;y)-14 y^3 (1+y)^2
\, H(0,0,0,0 ;y)\nonumber \\
& & {}+\frac{1}{216}  \left(71-24
\left(-71+16 \pi ^2\right) y-2704 y^2+2704 y^6-24 \left(71+16 \pi ^2\right)
y^7-71 y^8 \right. \nonumber \\
& & {}   +24 y^3 \left(-601+124 \pi ^2-288 \, \zeta(3)\right)+24 y^5
\left(601+124 \pi ^2-288 \, \zeta(3)\right) \nonumber \\
& & {} \left.+96 y^4 \left(19 \pi ^2-144
\, \zeta(3)\right)\right)\, H(-1 ;y)+\frac{1}{216} y  \left(1269-2550
y+8266 y^5 \right. \nonumber \\
& & {}  -5547 y^6-284 y^7-3 \pi ^2 \left(9-453 y^2-718 y^3-453 y^4+9
y^6\right) \nonumber \\
& & {} +6 y^3 (3535+1152 \, \zeta(3))+y^2 (-15937+3456 \, \zeta(3)) \nonumber \\
& & {} \left. +y^4
(41759+3456 \, \zeta(3))\right)\, H(0 ;y) \nonumber \\
& & {} +\frac{1}{2592}\left(3115-105152 y^2+105152
y^6-3115 y^8-1368 \pi ^4 y^3 (1+y)^2 \right. \nonumber \\
& & {}  +3 \pi ^2 \left(-21-1584 y+1740 y^2+20088
y^3+2352 y^4-11448 y^5-4684 y^6 \right. \nonumber \\
& & {} \left.+1920 y^7+125 y^8\right)+y (74760-7776
\, \zeta(3))+268704 y^4 \, \zeta(3) \nonumber \\
& & {}  -24 y^7 (3115+324 \, \zeta(3))+24
y^3 (-25145+7992 \, \zeta(3)) \nonumber \\
& & {}  \left. +24 y^5 (25145+7992 \, \zeta(3))\right)
\bigg\}.
\end{eqnarray}
\begin{equation}
T_3( q^2, y ,\mu, \epsilon ) = C^3( \epsilon ) \left( \frac{ \mu^2 }{
q^2 } \right)^{3 \epsilon} \frac{ (q^2)^3 }{ 2^{11} \pi^5 (1 + y)^8 } \, \cT_3 (
y,
\epsilon )\,,
\end{equation}
with
\begin{eqnarray}
 \cT_3 ( y, \epsilon ) & = & 
%
%
-\frac{1}{72} (1+y) \left(-3-109 y+\left(-443+24 \pi ^2\right) y^2+\left(-277+72
\pi ^2\right) y^3 \right. \nonumber \\
& & {} \left. +\left(277+72 \pi ^2\right) y^4+\left(443+24 \pi ^2\right)
y^5+109 y^6+3 y^7\right)\nonumber \\
& & {}+\frac{1}{3} y \left(2+18 y+42 y^2+57 y^3+42 y^4+18
y^5+2 y^6\right) \, H(0 ;y)\nonumber \\
& & {}-4 y^2 (1+y)^4 \, H(-1,0 ;y)+2 y^2
(1+y)^4 \, H(0,0 ;y) 
\nonumber \\
%
%
& & {} + \epsilon \bigg\{ 4 y^2 (1+y)^4 \, H(-3 ;y) \nonumber \\
& & {} +\frac{2}{3} y
\left(2+18 y+42 y^2+57 y^3+42 y^4+18 y^5+2 y^6\right)
\, H(-2 ;y)\nonumber \\
& & {}+\frac{1}{36} \left(3+112 y+24 \left(23+4 \pi ^2\right)
y^2+48 \left(15+8 \pi ^2\right) y^3+576 \pi ^2 y^4 \right. \nonumber \\
& & {}\left. +48 \left(-15+8 \pi ^2\right)
y^5+24 \left(-23+4 \pi ^2\right) y^6-112 y^7-3 y^8\right)
\, H(-1 ;y)\nonumber \\
& & {}-\frac{1}{18} y \left(-90+3 \left(-235+8 \pi ^2\right) y+6
\left(-189+16 \pi ^2\right) y^2+\left(-653+144 \pi ^2\right) y^3 \right. \nonumber \\
& & {} \left. +6 \left(51+16
\pi ^2\right) y^4+3 \left(133+8 \pi ^2\right) y^5+134 y^6+6 y^7\right)
\, H(0 ;y)\nonumber \\
& & {}-\frac{5}{36} \left(-3-112 y-552 y^2-720 y^3+720 y^5+552
y^6+112 y^7+3 y^8\right) \, H(1 ;y)\nonumber \\
& & {}+\frac{10}{3} y \left(2+18 y+42
y^2+57 y^3+42 y^4+18 y^5+2 y^6\right) \, H(2 ;y)\nonumber \\
& & {}+20 y^2 (1+y)^4
\, H(3 ;y)+28 y^2 (1+y)^4 \, H(-2,0 ;y)\nonumber \\
& & {}-8 y^2 (1+y)^4
\, H(-1,-2 ;y)\nonumber \\
& & {}+\frac{4}{3} y \left(5+33 y+63 y^2+80 y^3+63 y^4+33 y^5+5
y^6\right) \, H(-1,0 ;y)\nonumber \\
& & {}-40 y^2 (1+y)^4
\, H(-1,2 ;y)-\frac{2}{3} y (1+y)^4 \left(1-7 y+y^2\right)
\, H(0,0 ;y)\nonumber \\
& & {}-56 y^2 (1+y)^4 \, H(-1,-1,0 ;y)+12 y^2 (1+y)^4
\, H(-1,0,0 ;y)\nonumber \\
& & {}-6 y^2 (1+y)^4 \, H(0,0,0 ;y)\nonumber \\
& & {}+\frac{1}{432}
\left(213-8 \left(-985+36 \pi ^2\right) y-8 \left(985+36 \pi ^2\right) y^7-213
y^8 \right. \nonumber \\
& & {}\left.-36 y^2 \left(-1067+88 \pi ^2+120 \, \zeta(3)\right)-36 y^6 \left(1067+88
\pi ^2+120 \, \zeta(3)\right) \right. \nonumber \\
& & {}  -24 y^3 \left(-2077+336 \pi ^2+720
\, \zeta(3)\right)-24 y^5 \left(2077+336 \pi ^2+720 \, \zeta(3)\right)\nonumber \\
& & {}\left.-24
y^4 \left(467 \pi ^2+1080 \, \zeta(3)\right)\right) \bigg\}
\nonumber \\
%
%
& & {} + \epsilon ^2 \bigg\{ -12 y^2 (1+y)^4 \, H(-4 ;y)-\frac{4}{3} y
(1+y)^4 \left(1-7 y+y^2\right) \, H(-3 ;y)\nonumber \\
& & {}-\frac{1}{9} y \left(-90+15
\left(-47+16 \pi ^2\right) y+6 \left(-189+160 \pi ^2\right) y^2+\left(-653+1440
\pi ^2\right) y^3 \right. \nonumber \\
& & {}\left. +6 \left(51+160 \pi ^2\right) y^4+3 \left(133+80 \pi ^2\right)
y^5+134 y^6+6 y^7\right) \, H(-2 ;y)\nonumber \\
& & {}-\frac{5}{216} \left(-213-7880
y-38412 y^2-49848 y^3+49848 y^5+38412 y^6 \right. \nonumber \\
& & {} \left. +7880 y^7+213 y^8\right)
\, H(1 ;y)\nonumber \\
& & {}-\frac{5}{9} y \left(-90-705 y-1134 y^2-653 y^3+306 y^4+399
y^5+134 y^6+6 y^7\right) \, H(2 ;y)\nonumber \\
& & {}-\frac{20}{3} y (1+y)^4 \left(1-7
y+y^2\right) \, H(3 ;y)-60 y^2 (1+y)^4 \, H(4 ;y)\nonumber \\
& & {}+8 y^2
(1+y)^4 \, H(-3,-1 ;y)-132 y^2 (1+y)^4 \, H(-3,0 ;y)\nonumber \\
& & {}+40 y^2
(1+y)^4 \, H(-3,1 ;y)+56 y^2 (1+y)^4
\, H(-2,-2 ;y)\nonumber \\
& & {}+\frac{4}{3} y \left(2+18 y+42 y^2+57 y^3+42 y^4+18 y^5+2
y^6\right) \, H(-2,-1 ;y)\nonumber \\
& & {}-\frac{4}{3} y \left(19+87 y+105 y^2+104
y^3+105 y^4+87 y^5+19 y^6\right) \, H(-2,0 ;y)\nonumber \\
& & {}+\frac{20}{3} y
\left(2+18 y+42 y^2+57 y^3+42 y^4+18 y^5+2 y^6\right) \, H(-2,1 ;y)\nonumber \\
& & {}+280
y^2 (1+y)^4 \, H(-2,2 ;y)+24 y^2 (1+y)^4
\, H(-1,-3 ;y)\nonumber \\
& & {}+\frac{8}{3} y \left(5+33 y+63 y^2+80 y^3+63 y^4+33 y^5+5
y^6\right) \, H(-1,-2 ;y)\nonumber \\
& & {}-\frac{1}{18} (1+y) \left(-3-109
y-\left(443+960 \pi ^2\right) y^2-\left(277+2880 \pi ^2\right)
y^3 \right. \nonumber \\
& & {} \left. +\left(277-2880 \pi ^2\right) y^4+\left(443-960 \pi ^2\right) y^5+109 y^6+3
y^7\right) \, H(-1,-1 ;y)\nonumber \\
& & {}+\frac{1}{18} \left(-21+116 y-6 \left(-415+7
\pi ^2\right) y^2+\left(4740-168 \pi ^2\right) y^3 \right. \nonumber \\
& & {} +\left(5018-252 \pi ^2\right)
y^4+\left(1860-168 \pi ^2\right) y^5-6 \left(-47+7 \pi ^2\right) y^6 \nonumber \\
& & {} \left.-332 y^7-33
y^8\right) \, H(-1,0 ;y)\nonumber \\
& & {}-\frac{5}{18} \left(-3-112 y-552 y^2-720
y^3+720 y^5+552 y^6+112 y^7+3 y^8\right) \, H(-1,1 ;y)\nonumber \\
& & {}+\frac{40}{3} y
\left(5+33 y+63 y^2+80 y^3+63 y^4+33 y^5+5 y^6\right) \, H(-1,2 ;y)\nonumber \\
& & {}+120
y^2 (1+y)^4 \, H(-1,3 ;y)\nonumber \\
& & {}+\frac{1}{18} y (1+y) \left(-90+3 \left(-89+7
\pi ^2\right) y+\left(-87+63 \pi ^2\right) y^2+\left(190+63 \pi ^2\right)
y^3 \right.  \nonumber \\
& & {} \left. +\left(176+21 \pi ^2\right) y^4+19 y^5+3 y^6\right)
\, H(0,0 ;y)\nonumber \\
& & {}-\frac{5}{18} \left(-3-112 y-552 y^2-720 y^3+720 y^5+552
y^6+112 y^7+3 y^8\right) \, H(1,-1 ;y)\nonumber \\
& & {}-\frac{5}{9} \left(-3-112 y-552
y^2-720 y^3+720 y^5+552 y^6+112 y^7+3 y^8\right)
\, H(1,0 ;y)\nonumber \\
& & {}-\frac{25}{18} \left(-3-112 y-552 y^2-720 y^3+720 y^5+552
y^6+112 y^7+3 y^8\right) \, H(1,1 ;y)\nonumber \\
& & {}+\frac{20}{3} y \left(2+18 y+42
y^2+57 y^3+42 y^4+18 y^5+2 y^6\right) \, H(2,-1 ;y)\nonumber \\
& & {}+\frac{40}{3} y
\left(2+18 y+42 y^2+57 y^3+42 y^4+18 y^5+2 y^6\right)
\, H(2,0 ;y)\nonumber \\
& & {}+\frac{100}{3} y \left(2+18 y+42 y^2+57 y^3+42 y^4+18 y^5+2
y^6\right) \, H(2,1 ;y)\nonumber \\
& & {}+40 y^2 (1+y)^4 \, H(3,-1 ;y)+80 y^2
(1+y)^4 \, H(3,0 ;y)\nonumber \\
& & {}+200 y^2 (1+y)^4 \, H(3,1 ;y)+296 y^2
(1+y)^4 \, H(-2,-1,0 ;y)\nonumber \\
& & {}-36 y^2 (1+y)^4 \, H(-2,0,0 ;y)-16 y^2
(1+y)^4 \, H(-1,-2,-1 ;y)\nonumber \\
& & {}+264 y^2 (1+y)^4 \, H(-1,-2,0 ;y)-80
y^2 (1+y)^4 \, H(-1,-2,1 ;y)\nonumber \\
& & {}-112 y^2 (1+y)^4
\, H(-1,-1,-2 ;y)\nonumber \\
& & {}+\frac{8}{3} y \left(23+123 y+189 y^2+218 y^3+189
y^4+123 y^5+23 y^6\right) \, H(-1,-1,0 ;y)\nonumber \\
& & {}-560 y^2 (1+y)^4
\, H(-1,-1,2 ;y)\nonumber \\
& & {}-4 y (1+y)^4 \left(1-7 y+y^2\right)
\, H(-1,0,0 ;y)\nonumber \\
& & {}-80 y^2 (1+y)^4 \, H(-1,2,-1 ;y)-160 y^2
(1+y)^4 \, H(-1,2,0 ;y)\nonumber \\
& & {}-400 y^2 (1+y)^4
\, H(-1,2,1 ;y)\nonumber \\
& & {}+\frac{2}{3} y \left(1-27 y-105 y^2-159 y^3-105 y^4-27
y^5+y^6\right) \, H(0,0,0 ;y)\nonumber \\
& & {}-592 y^2 (1+y)^4
\, H(-1,-1,-1,0 ;y)+72 y^2 (1+y)^4 \, H(-1,-1,0,0 ;y)\nonumber \\
& & {}-28 y^2
(1+y)^4 \, H(-1,0,0,0 ;y)+14 y^2 (1+y)^4
\, H(0,0,0,0 ;y)\nonumber \\
& & {}-\frac{1}{108} y  \left(-2556+13204
y^6+426 y^7+3 \pi ^2 \left(18+354 y+1050 y^2+1513 y^3 \right. \right. \nonumber \\
& & {} \left. +1050 y^4+354 y^5+18
y^6\right)+27 y^5 (2257+64 \, \zeta(3))+96 y^4 (1021+72 \, \zeta(3))\nonumber \\
& & {}\left. +48
y^2 (-35+144 \, \zeta(3))+9 y (-1765+192 \, \zeta(3))+y^3 (58955+10368
\, \zeta(3))\right)\, H(0 ;y)\nonumber \\
& & {}+\frac{1}{216}  \left(213-8
\left(-985+192 \pi ^2\right) y-8 \left(985+192 \pi ^2\right) y^7-213 y^8 \right. \nonumber \\
& & {}  -24 y^5
\left(2077+1008 \pi ^2-1152 \, \zeta(3)\right)-96 y^4 \left(331 \pi ^2-432
\, \zeta(3)\right)\nonumber \\
& & {} -36 y^2 \left(-1067+320 \pi ^2-192 \, \zeta(3)\right)-36
y^6 \left(1067+320 \pi ^2-192 \, \zeta(3)\right)\nonumber \\
& & {}\left. +y^3 \left(49848-24192 \pi
^2+27648 \, \zeta(3)\right)\right)\, H(-1 ;y)\nonumber \\
& & {}+\frac{1}{2592}\left(9345-9345 y^8+1368 \pi
^4 y^2 (1+y)^4+3 \pi ^2 \left(-63-6672 y \right. \right. \nonumber \\
& & {}  \left. -48216 y^2-75792 y^3-37392 y^4+29328
y^5+32376 y^6+9680 y^7+375 y^8\right)\nonumber \\
& & {}  -1426464 y^4 \, \zeta(3)-324 y^2
(-5123+1184 \, \zeta(3))-324 y^6 (5123+1184 \, \zeta(3))\nonumber \\
& & {} -8 y (-42917+3888
\, \zeta(3)) -8 y^7 (42917+3888 \, \zeta(3))\nonumber \\
& & {}\left. -24 y^3 (-89405+42336
\, \zeta(3))-24 y^5 (89405+42336 \, \zeta(3))\right) \bigg\}.
\end{eqnarray}
\begin{equation}
T_4( q^2, y ,\mu, \epsilon ) = C^3( \epsilon ) \left( \frac{ \mu^2 }{
q^2 } \right)^{3 \epsilon} \frac{ q^2 }{ 2^{11} \pi^5 } \, \cT_4 ( y,
\epsilon )\,,
\end{equation}
with
\begin{eqnarray}
\cT_4 ( y, \epsilon ) & = &
\left[ \frac{3}{(1+y)^4}-\frac{6}{(1+y)^3}+\frac{4}{(1+y)^2}-\frac{1}{1+y}
\right] H( 0 ; y )
\nonumber \\[0.2cm]
& & {} + \left[ \frac{4}{(1+y)^3}-\frac{7}{(1+y)^2}+\frac{3}{1+y}
\right] H( 0,0 ; y )
\nonumber \\[0.2cm]
& & {} + \left[ \frac{2}{(1+y)^2}-\frac{2}{1+y}\right] H( -1,0 ; y)
\nonumber \\[0.2cm]
& & {} - \frac{1}{4}+\frac{3}{(1+y)^3}+\frac{-27+\pi ^2}{6 (1+y)^2}+\frac{12-\pi
^2}{6(1+y)}
\nonumber \\[0.2cm]
& & {} + \epsilon\, \bigg\{ \left[
\frac{8}{(1+y)^3}-\frac{14}{(1+y)^2}+\frac{6}{1+y} \right] H( -3 ;y )
\nonumber \\[0.2cm]
& & {} + \left[ \frac{6}{(1+y)^4}-\frac{12}{(1+y)^3}+\frac{8}{(1+y)^2
}-\frac{2}{1+y}\right] H( -2 ; y)
\nonumber \\[0.2cm]
& & {} + \left[ -\frac{1}{2}+\frac{6}{(1+y)^3}+\frac{-27-4 \pi ^2}{3
(1+y)^2}+\frac{4 \left(3+\pi ^2\right)}{3 (1+y)}\right]
H( -1 ; y)
\nonumber \\[0.2cm]
& & {} + \left[ -2+\frac{39}{2 (1+y)^4}+\frac{-27-4 \pi
^2}{(1+y)^3}+\frac{9+20 \pi ^2}{3 (1+y)^2}+\frac{39-16 \pi ^2}{6 (1+y)}\right]
H( 0 ; y)
\nonumber \\[0.2cm]
& & {} + \left[ -\frac{5}{2}+\frac{30}{(1+y)^3}-\frac{45}{(1+y)^2}
+\frac{20}{1+y}\right] H( 1 ; y)
\nonumber \\[0.2cm]
& & {} + \left[ \frac{30}{(1+y)^4}-\frac{60}{(1+y)^3}+\frac{40}{
(1+y)^2}-\frac{10}{1+y}\right] H( 2 ; y)
\nonumber \\[0.2cm]
& & {} +\left[ \frac{40}{(1+y)^3}-\frac{70}{(1+y)^2}+\frac{30}{1+y}
\right] H( 3; y)
\nonumber \\[0.2cm]
& & {} + \left[ \frac{40}{(1+y)^3}-\frac{74}{(1+y)^2}+\frac{34} {1+y}
\right] H( -2,0; y)
\nonumber \\[0.2cm]
& & {} + \left[ \frac{4}{(1+y)^2}-\frac{4}{1+y}\right]
H( -1,-2; y)
\nonumber \\[0.2cm]
& & {} +\left[ \frac{24}{(1+y)^4}-\frac{48}{(1+y)^3}+\frac{43}{
(1+y)^2}-\frac{19}{1+y}\right] H( -1,0; y)
\nonumber \\[0.2cm]
& & {} + \left[ \frac{20}{(1+y)^2}-\frac{20}{1+y}\right]
H( -1,2; y)
\nonumber \\[0.2cm]
& & {} +\left[ \frac{18}{(1+y)^3}-\frac{65}{2 (1+y)^2}+\frac{29}{2
(1+y)}\right] H( 0,0; y )
\nonumber \\[0.2cm]
& & {} +\left[ \frac{28}{(1+y)^2}-\frac{28}{1+y}
\right] H( -1,-1,0; y )
\nonumber \\[0.2cm]
& & {} + \left[ \frac{24}{(1+y)^3}-\frac{42}{(1+y)^2}+\frac{18}{
1+y}\right] H( -1,0,0; y)
\nonumber \\[0.2cm]
& & {} +\left[-\frac{12}{(1+y)^3}+\frac{21}{(1+y)^2}-\frac{9}{
1+y}\right] H( 0,0,0; y)
\nonumber \\[0.2cm]
& & {} + \left[ \frac{8}{(1+y)^3}-\frac{12}{(1+y)^2}+\frac{4}{1+y}
\right] H( 1,0,0; y)
\nonumber \\[0.2cm]
& & {} - \frac{25}{8}-\frac{7 \pi ^2}{2
(1+y)^4}+\frac{100+\pi ^2-44 \zeta( 3 ) }{4
(1+y)}
\nonumber \\[0.2cm]
& & {} + \frac{75+14 \pi ^2-24 \zeta( 3 ) }{2 (1+y)^3}+\frac{-225-15 \pi
^2+92 \zeta( 3 ) }{4 (1+y)^2} \bigg\}.
\end{eqnarray}
\begin{equation}
T_5( q^2, y ,\mu, \epsilon ) = C^3( \epsilon ) \left( \frac{ \mu^2 }{
q^2 } \right)^{3 \epsilon} \frac{ 1 }{ 2^{11} \pi^5 } \, \cT_5 ( y,
\epsilon )\,,
\end{equation}
with
\begin{eqnarray}
%
%
\cT_5 ( y, \epsilon ) & = & \left(\frac{2}{(1+y)^2}-\frac{2}{1+y}\right)
\, H ( 0; y ) - \left( 2 + \frac{4}{(1+y)^2} - \frac{4}{1+y}\right)
\, H ( -1,0; y )
\nonumber \\[0.2cm]
& & {} + \left(4+\frac{6}{(1+y)^2}-\frac{10}{1+y}\right)
\, H( 0,0; y ) + 
\left(2+\frac{8}{(1+y)^2}-\frac{8}{1+y}\right) \, H( 1,0; y )
\nonumber \\[0.2cm]
& & {} + 1 + \frac{\pi^2}{6} +\frac{\pi
^2}{(1+y)^2} - \frac{ 2 + \pi^2}{1+y} 
\nonumber \\[0.2cm]
%
%
& & {} + \epsilon \, \bigg\{ 
\left(8+\frac{12}{(1+y)^2}-\frac{20}{1+y}\right)
\, H( -3; y ) + \left(\frac{4}{(1+y)^2}-\frac{4}{1+y}\right)
\, H( -2; y )
\nonumber \\[0.2cm]
& & {} + \left( 2 + \frac{10 \pi^2}{3} + \frac{32 \pi
^2}{3 (1+y)^2}-\frac{4 \left(3+8 \pi ^2\right)}{3 (1+y)}\right)
\, H( -1; y )
\nonumber \\[0.2cm]
& & {} + \left(8 - 4 \pi^2 + \frac{4 \left( 18 - 5 \pi
^2\right)}{3 (1+y)^2} - \frac{32 \left(3 - \pi ^2\right)}{3 (1+y)}\right)
\, H( 0; y )
\nonumber \\[0.2cm]
& & {} + \left( 10 -\frac{2 \pi^2}{3} -\frac{8 \pi ^2}{3
(1+y)^2}- \frac{4 \left(15-2 \pi ^2\right)}{3 (1+y)}\right)
\, H( 1; y)
\nonumber \\[0.2cm]
& & {} +\left(\frac{20}{(1+y)^2}-\frac{20}{1+y}\right)
\, H( 2; y ) + \left(40+\frac{60}{(1+y)^2}-\frac{100}{1+y}\right)
\, H( 3; y )
\nonumber \\[0.2cm]
& & {} + \left(40+\frac{52}{(1+y)^2}-\frac{92}{1+y}\right)
\, H( -2,0; y )- \left(4 + \frac{8}{(1+y)^2} - \frac{8}{1+y}\right)
\, H( -1,-2; y )
\nonumber \\[0.2cm]
& & {} - \left( 14 + \frac{8}{(1+y)^2} - \frac{8}{1+y}\right)
\, H( -1,0; y ) - \left(20 + \frac{40}{(1+y)^2} - \frac{40}{1+y}\right)
\, H( -1,2; y )
\nonumber \\[0.2cm]
& & {} + \left(28+\frac{40}{(1+y)^2}-\frac{68}{1+y}\right)
\, H( 0,0; y ) + \left(4+\frac{16}{(1+y)^2}-\frac{16}{1+y}\right)
\, H( 1,-2; y )
\nonumber \\[0.2cm]
& & {} + \left(14+\frac{56}{(1+y)^2}-\frac{56}{1+y}\right)
\, H( 1,0; y ) + \left(20+\frac{80}{(1+y)^2}-\frac{80}{1+y}\right)
\, H( 1,2; y )
\nonumber \\[0.2cm]
& & {} - \left( 28 + \frac{56}{(1+y)^2} - \frac{56}{1+y}\right)
\, H( -1,-1,0; y )
\nonumber \\[0.2cm]
& & {} + \left(24+\frac{36}{(1+y)^2}-\frac{60}{1+y}\right)
\, H( -1,0,0; y)
\nonumber \\[0.2cm]
& & {} + \left(12+\frac{48}{(1+y)^2}-\frac{48}{1+y}\right)
\, H( -1,1,0; y )
\nonumber \\[0.2cm]
& & {} - \left( 12 + \frac{18}{(1+y)^2} - \frac{30}{1+y}\right)
\, H( 0,0,0; y )
\nonumber \\[0.2cm]
& & {} + \left(12+\frac{48}{(1+y)^2}-\frac{48}{1+y}\right)
\, H( 1,-1,0; y)
\nonumber \\[0.2cm]
& & {} + \left(12+\frac{32}{(1+y)^2}-\frac{40}{1+y}\right)
\, H( 1,0,0; y )
\nonumber \\[0.2cm]
& & {} + \left(12+\frac{48}{(1+y)^2}-\frac{48}{1+y}\right)
\, H( 1,1,0; y )
\nonumber \\[0.2cm]
& & {} + \frac{1}{6} \left(84+7 \pi ^2-18
\zeta( 3 ) \right) + \frac{-28-5 \pi ^2-10 \zeta( 3 ) }{1+y}+\frac{5 \pi
^2+22 \zeta( 3 )}{(1+y)^2} \bigg\} + \cO( \epsilon^2 )\,.
\end{eqnarray}